\documentclass[aps,prl,amsmath,amssymb,reprint,superscriptaddress]{revtex4-2}
\usepackage{graphicx}
\usepackage{dcolumn}
\usepackage{bm}
\usepackage{hyperref}
\usepackage{xcolor}

\setcitestyle{super}

\begin{document}
	
	\preprint{APS/123-QED}
	
\title{Fluid Dynamical Pathways of Airborne Transmission while Waiting in a Line}

	\author{Ruixi Lou}
	\affiliation{Department of Physics, University of Massachusetts Amherst, MA 01003, USA.
	}
    \affiliation{Department of Physics, University of Chicago, Chicago, IL 60637, USA.
	}

 \author{Milo van Mooy}
\affiliation{Department of Physics, University of Massachusetts Amherst, MA 01003, USA.
	}

      \author{Gabriel A. Tarditti
}
	\affiliation{School of Engineering, University of Cadiz, Cádiz, Andalusia 11001, Spain.
	}
	
	\author{Rodolfo Ostilla Monico}
    \affiliation{School of Engineering, University of Cadiz, Cádiz, Andalusia 11001, Spain.
	}
    
	\author{Varghese Mathai\textsuperscript{*}}
	\affiliation{Department of Physics, University of Massachusetts Amherst, MA 01003, USA.
	}
	
\affiliation{Department of Mechanical \& Industrial Engineering, University of Massachusetts Amherst, MA 01003, USA\\ \noindent
\textsuperscript{*}\textnormal{Corresponding author: vmathai@umass.edu}
}

	
	\begin{abstract}
		\noindent 
		
Waiting in a line (or a queue) is an important, often unavoidable social interaction that occurs frequently in public spaces. Despite its wide prevalence and rich parametric variability, few studies have addressed the risks of airborne transmission while waiting in a line.  Here we use a combination of scaled-down laboratory experiments and direct numerical simulations (DNS) 
to assess the flow patterns and infection risks in a simplified waiting line setting.  We observed the presence of fluid dynamical counter-currents  -- due to the competing effects of line kinematics and thermal gradients --  which can either heighten or suppress the risks of transmission. 
Depending on the walking speed, an intermediate ambient temperature range can potentially heighten the infection risks by allowing the breath plume to linger in the air for extended durations; however, colder and warmer ambients both suppress the spread. The current guideline of increasing physical separation has limited impact on reducing transmission in the waiting line setting. The present work highlights the need for updated transmission mitigation guidelines that go beyond the simplicity of the ``six-feet rule'' in social interactions where physical separation, duration of interaction, and periodicity of movements are factors.

	\end{abstract}

	\maketitle
	\newpage

The COVID-19 pandemic has redefined the ways in which we interact in social settings. From social distancing and mask-wearing to testing and contact tracing, these measures have proven to be effective in mitigating the ever-present risks of infections. 
It has now been recognized that a variety of respiratory pathogens, including RSV, SARS-CoV-2, and Influenza are spread primarily through tiny droplets and airborne particles released by asymptomatic and presymptomatic individuals during speaking and breathing \cite{bourouiba2021fluid}. These respiratory contaminants are typically less than 10 microns in size \cite{abkarian2020speech, pohlker2021respiratory}, which can enable them to remain suspended in the air for minutes to several hours depending on the ambient conditions \cite{chong2021extended,bourouiba2020turbulent,mittal2023flow}.  
Although physical distancing and the wearing of face masks are common measures used to reduce the spread of infections \cite{Tang2011,Lai2012,Greenhalgh2020,Leung2020,lee2008respiratory,plasticpanel2020roshun,Morawska2020,zhang2020identifying,yu2004evidence,abkarian2020stretching}, these provide little insight into the transport of aerosols in situations where the duration of interactions and the time-history of the movements of individuals becomes a factor.  The problem is further intensified under varying influences of thermally gradients and buoyancy-driven plumes in indoor spaces, which can vary significantly across climate zones and biomes.  \cite{he2021airborne,bhagat2020effects,Gupta2010,bourouiba2020turbulent,Meselson2020,yan2018infectious,wolfel2020virological,yang2011concentrations,scharfman2016visualization,bahl2020airborne,bourouiba2014violent,chong2021extended,van2020aerosol,stadnytskyi2020airborne,mathai2022aerosol,yang2020towards,beggs2010potential}.

Almost all of the studies of airborne transmission take the perspective of infections occurring under ``static'' interaction conditions, such as across desks in a classroom setting or individuals interacting in a face-to-face meeting standing or while seated in a passenger car next to an infected individual \cite{mathai2021air,abkarian2020speech}.   Consequently, current guidelines for transmission mitigation are based almost entirely on static, physical distancing-based recommendations -- such as the ``six-feet'' rule and limiting the duration of interactions. In contrast, a wide class of social interactions occur in our daily lives that are distinctly outside the domain of  ``static'' interactions. A familiar example is that of waiting and walking in a line -- at a grocery store or a vaccination clinic or an airport security. The indoor waiting line introduces considerable complexity to the modeling of aerosol transmission due to the additional time and length scales of periodic walking and stopping, the physical separation (guided or self-imposed), and the variations in ambient temperature. 
In particular, currents generated by the movement of people interact with the aerosols, advecting, and diffusing, while also being convected (upward or downward) by the thermal gradient effects with the ambient \cite{tang2009schlieren, mittal2020flow, mittal2023flow} can lead to complicated pathways of infection. Despite the ubiquity of these interactions, few studies have addressed airborne cross-infection risks in a waiting line (queue). 

 \begin{figure*}[!htbp]
\centering
\includegraphics[width=0.96\linewidth]{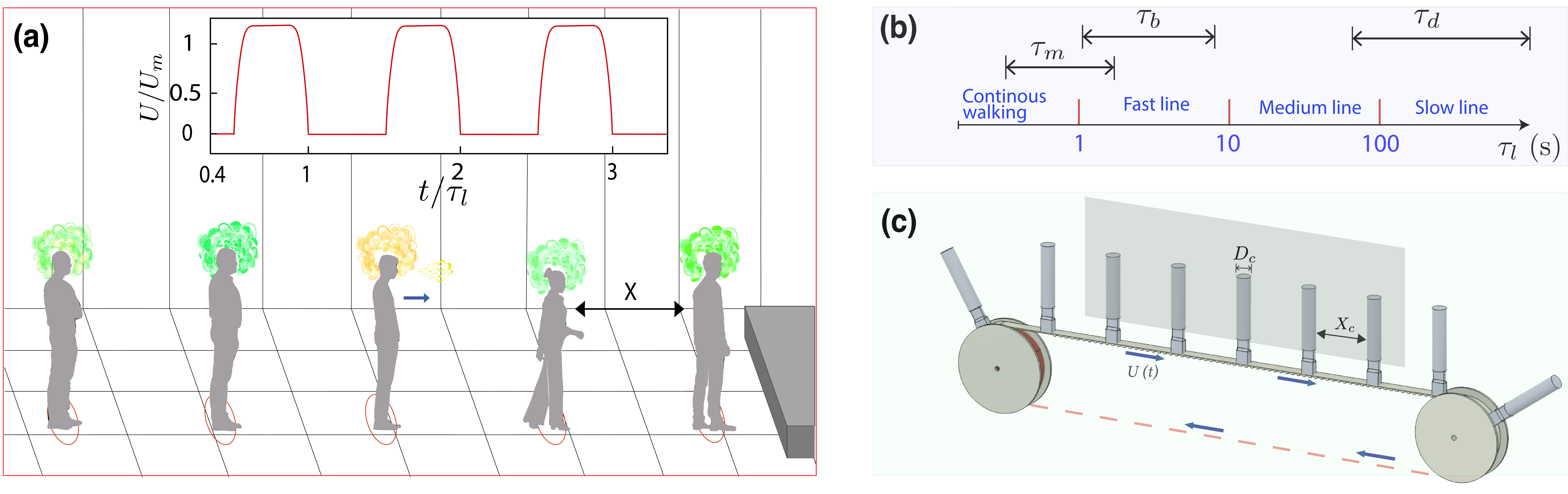}
\caption{\footnotesize{Representations of airborne transmission while waiting in a line. (a) A simplified schematic representation of people waiting and waking in a line. The green clouds represent the ``breath plumes'' of individuals, while the yellow cloud represents the breath plume released by a potentially infectious individual in the line. The inset shows an example of the velocity, which is marked by periodic start-stop cycles, where $U(t)$ is the instantaneous velocity, $U_m$ is the walking speed, and $\tau_l$ is the time period of the line. (b) Comparison of the moving time-scale $\tau_m$ with the characteristic time-scales of diffusion $\tau_d$ and buoyancy $\tau_b$. We classify the line movement as fast/medium/slow depending on the time period $\tau_l$ of the waiting line movement. Note that the duration of the walking, $\tau_m$, is much shorter than $\tau_l$, except for a continuously moving (or fast) line. (c) To mimic the line kinematics we employ a pulley-driven conveyor belt system carrying equally spaced cylindrical dummies on a belt, as shown by the conveyor-belt schematic. Here, $D_c$ is the diameter of the cylinder, $X_c$ is the separation between cylinders, and 17 cylindrical dummies are installed over the whole belt, with roughly 7 cylinders above (below) the pulleys. The three centrally-located cylinders in the gray-shaded area represent the region of interest in experiments.}}
\label{figure1_schematic}
\vspace{-0.2 cm}
\end{figure*}

Here we study the simplified case of periodically walking and halting individuals in a waiting line, by using a combination of laboratory experiments and direct numerical simulations (DNS). The movements of people in the waiting area can be separated into two stages: a waiting period, followed by a brief walking period (Fig.~\ref{figure1_schematic}a). We ask the question: {\it where do the aerosols released by an infectious individual in a waiting line end up (denoted by the yellow cloud in Fig.~\ref{figure1_schematic}a)?} During the waiting phase, we assume that an infectious person continually releases (breathes) airborne particles around them which slowly diffuse into the surrounding air. After some time elapses, they walk to the next position in line. The resulting reorganization of individuals advects and diffuses the airborne particles. 
Except in very slow situations, the diffusion may be considered sub-dominant (Fig~\ref{figure1_schematic}b), with the characteristic diffusion time-scale $\tau_d$ much larger than the average waiting period $\tau_l$. Two distinct processes may disperse the virus-laden aerosols: (i) the flows (air currents) induced by walking humans and (ii) the buoyancy (thermally driven) of breath plumes. Remarkably, for a variety of waiting-line scenarios, the time scales of walking and buoyancy can be comparable, and a complex, coupled interplay between the two can be expected.
We will reveal how the competition between shear (line kinematics) and buoyancy (thermal gradients) emerges as an important criterion for transmission risks in the waiting line. 

 \begin{figure*}[!tbp]
\centering
\includegraphics[width=1.0\linewidth]{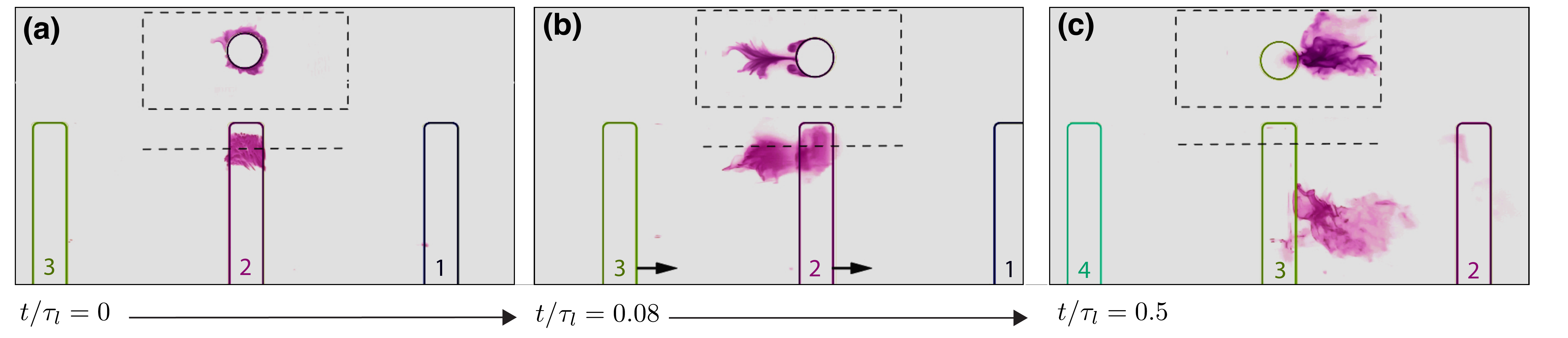}
\caption{\footnotesize{Experimental visualization of the transport of released (in pink) dye during a walk-waiting cycle in a laboratory-scale realization of a waiting line. (a) Distribution of the fluorescein disodium dye in side view and top view at $t/\tau_l=0$ (the moment that the cylinders start to move), (b) $t/\tau_l=0.08$ (the cylinders are moving) and (c) $t/\tau_l=0.5$ ($\approx 0.33 \ t/\tau_l$ after the cylinders stop). Soon after the cylinders stop, the dye suddenly seemingly sinks toward the ground. Note that the dye is perfectly density-matched with the fluid. Here, 1, 2, 3, and 4 are numbers dummy, and $\tau_m/\tau_l \approx 0.16$ in the above example.}}
\label{figure2_dye_DNS_currentsettling}
\vspace{-0.2 cm}
\end{figure*}

A number of practical challenges exist to conducting field experiments with human subjects in a waiting line, while also varying the physical parameters in the problem. Hence, we apply the concept of dynamic similarity to transform the waiting line to a laboratory setting with a controlled, driven conveyor-belt system of cylindrical dummies (Fig.~\ref{figure1_schematic}c and see SI~\cite{supp}).  In combination with the experiments, fully resolved direct numerical simulations (DNS) were employed to gain further insights into the emergent flow patterns and the Lagrangian trajectories of the tracer particles.  The problem combines the effects of unsteady kinematics with buoyancy \cite{niemela2000turbulent,ahlers2009heat,iyer2020classical,lohse2010small,xia2013current,chaigne2023dissolution}; hence, these may be reduced into a set of non-dimensional parameters that characterize the flow: a Reynolds number, a Grashof number, a reduced frequency, and a dimensionless spacing. The Reynolds number is defined as $Re = U D_s/\nu$, where $D_s$ and $U$ are the characteristic length and velocity scales, respectively, and $\nu$ is the kinematic viscosity of the fluid. For the waiting line, the relevant characteristic length is $D_s$, the shoulder width of a person, and $U$ is a characteristic walking speed, taken as the walking speed $U_m$. While the average unimpeded walking speed of humans is around 3.3 mph ($\sim 1.5$ m/s), typically in indoor spaces and under recurrent start-stop conditions (such as in a waiting line), the peak walking speed may lie in the range of $0.5-0.87$ m/s \cite{wang2014tracking}. Consequently, for the typical waiting line, $Re$ can lie in the range $[1 -1.8] \times 10^{4}$, indicating turbulent flow.  For this range of $Re$, a nearly Reynolds-independent large-scale flow can be expected, and therefore, the shear--buoyancy competition dominates over any enhanced spreading/dilution of aerosols due to turbulent mixing. The strength of buoyancy effects coming from breath plumes is quantified through the Grashof number, $Gr ={g\beta \Delta T D_s^{3}}/{\nu^2}=U_b^2 D_s^2 / \nu^2$,  where $g$ is the gravitational acceleration, $\beta$ is the thermal expansion coefficient of the fluid,  $\Delta T$ is the temperature difference between human exhalation and the ambient background, and $U_b=\sqrt{g\beta\Delta T D_s}$ is the characteristic free-fall (buoyancy) velocity. We may note that a typical temperature difference of $\Delta T=4^\circ C$ corresponds to $Gr\approx2.5\times 10^7$. In addition, we define the velocity ratio $\hat{U}=U_b/U_m = \sqrt{Gr}/Re$ as a way of measuring the strength of buoyancy compared to advection. The two other non-dimensional parameters in the system are the non-dimensional separation, defined as $A^*=X/D_s$, where $X$ is the separation between people in the waiting line, and the non-dimensional frequency of motion, defined as $f^{*}=\tau_m/\tau_l$, where $\tau_m=X/U_m$ is the moving time and $\tau_l$ is the total period comprised of moving and waiting times during a cycle of motion (i.e.~the time between starting to move from one spot to a new spot and again starting to move from that spot).

In the present study, we limit the dimensionless $A^*=5$ (equivalent to six-foot separation) and $f^*=0.17$, i.e. a waiting time that is five times longer than the walking time.  We varied $A^* \in [2.5-10]$ and $f^*$ covering the range of medium to slow lines (Fig.~\ref{figure1_schematic} from $\tau_l \sim 10-100$ s), it was evident that physical separation, surprisingly, has only a minor effect on aerosol spreading (see SI~\cite{supp} and Fig.~S2). 
The risk of cross-infection was highest for the nearby individuals in the line, as there was not sufficient time for the released airborne particles to disperse. Thus, transport from the infected individual to the trailing susceptible individual will be our focus in the present work.

{\bf Experiments:} We rely on a basic model of a waiting line composed of periodically moving circular cylinders to represent humans, as shown in the schematic of Figure \ref{figure1_schematic}c. The experiments were conducted in a laboratory scale, driven conveyor belt system of periodically moving cylindrical dummies. Although we had the option of 3D-printing scaled human-like dummies, we believe the primary flow features are sufficiently captured even under the simplification of the complex geometry to a cylindrical bluff body. The belt was driven by a stepper motor and controlled using an Arduino microcontroller. Since the kinematic viscosity $\nu$ of water is nearly one order of magnitude lower than that of air, we can scale down the physical size of the experiment while also ensuring geometric, kinematic, and dynamic similarity of the physical system \cite{donelan2000exploring}. For the in-lab conditions, the scaled height and width of the average person were 110 mm and 25 mm, respectively, and the recommended 2 m ($\approx$ 6 ft) separation between individuals was scaled down to 125 mm. The walking time, the walking speed, and the waiting time were precisely controlled using a stepper motor driving the conveyor belt system (Fig.~\ref{figure1_schematic}c and SI). All physical variables were scaled down (or up) to match the relevant dimensionless numbers. During the experiments, we used UV-induced fluorescence (see Fig.~\ref{figure2_dye_DNS_currentsettling} and SI), combined with time-resolved Particle Imaging Velocimetry (PIV, see SI~\cite{supp} for details) to visualize the passive/active scalar transport and the underlying fluid motion. 
To match buoyancy effects of dispersed phase with a density mismatch\cite{almeras2017experimental}, the temperature-density equivalence, applicable for small thermal gradients, was employed in the experiments. The temperature-density correspondence, when transforming from a waiting line in the air to a laboratory setting, is 

\begin{equation}           
  {\Delta \Gamma} =  \left (\displaystyle\frac{D_s}{D_c}\right )^3 \left( \displaystyle\frac{\nu}{\nu_a} \right)^2  \beta \Delta T,
\end{equation}
where $\Delta\Gamma$ is the proportional density increase, $\Delta\Gamma=\Delta \rho/\rho$,  where $\nu$  and $\nu_a$ are the kinematic viscosities of the fluid and that of air, respectively, and $D_c$ is the cylinder diameter. For reference, $\Delta \Gamma=0.16$ in our experiment would correspond to a temperature difference of about $4^\circ$C between breath and ambient, following a model for exhaled breath temperature as a function of ambient temperature and skin temperature (see Refs. \cite{chong2021extended,yang2022increased,houdas2013human} or SI~\cite{supp} for more details).

{\bf Numerical simulations:} Direct numerical simulations (DNS) were used to further validate the experimental observations and to gain deeper insights into particle transport. We simulated the (non-dimensional) incompressible Navier-Stokes equations:

\begin{equation}
     \frac{\partial \textbf{u}}{\partial t}+\textbf{u}\cdot \nabla \textbf{u}=-\nabla p + \frac{1}{Re} \nabla^2 \textbf{u} + \textbf{f}_b + \textbf{f}_i, 
     \label{eq:navierstokes}
\end{equation}

\noindent in a straight parallelepiped domain, where $\textbf{u}$ is the non-dimensional fluid velocity, $p$ the pressure and $t$ the time. Two body forces appear, a buoyancy force $\textbf{f}_b$ to model thermal effects under the Boussinesq approximation, and an immersed boundary force $\textbf{f}_i$, which incorporates the forcing on the fluid due to the movements of the cylinders \cite{fadlun2000combined}. We used the open-source code AFiD \cite{van2015pencil}, which has been widely used to simulate fluid dynamics problems with buoyancy and shear \cite{blass2020flow,blass2021flow,jin2022shear}. The code uses a centered, energy-conserving finite differences in space and a third-order Runge-Kutta fractional time-stepping.

\begin{figure}[!tbp]
 \centering
 \includegraphics[width=1.02\linewidth]{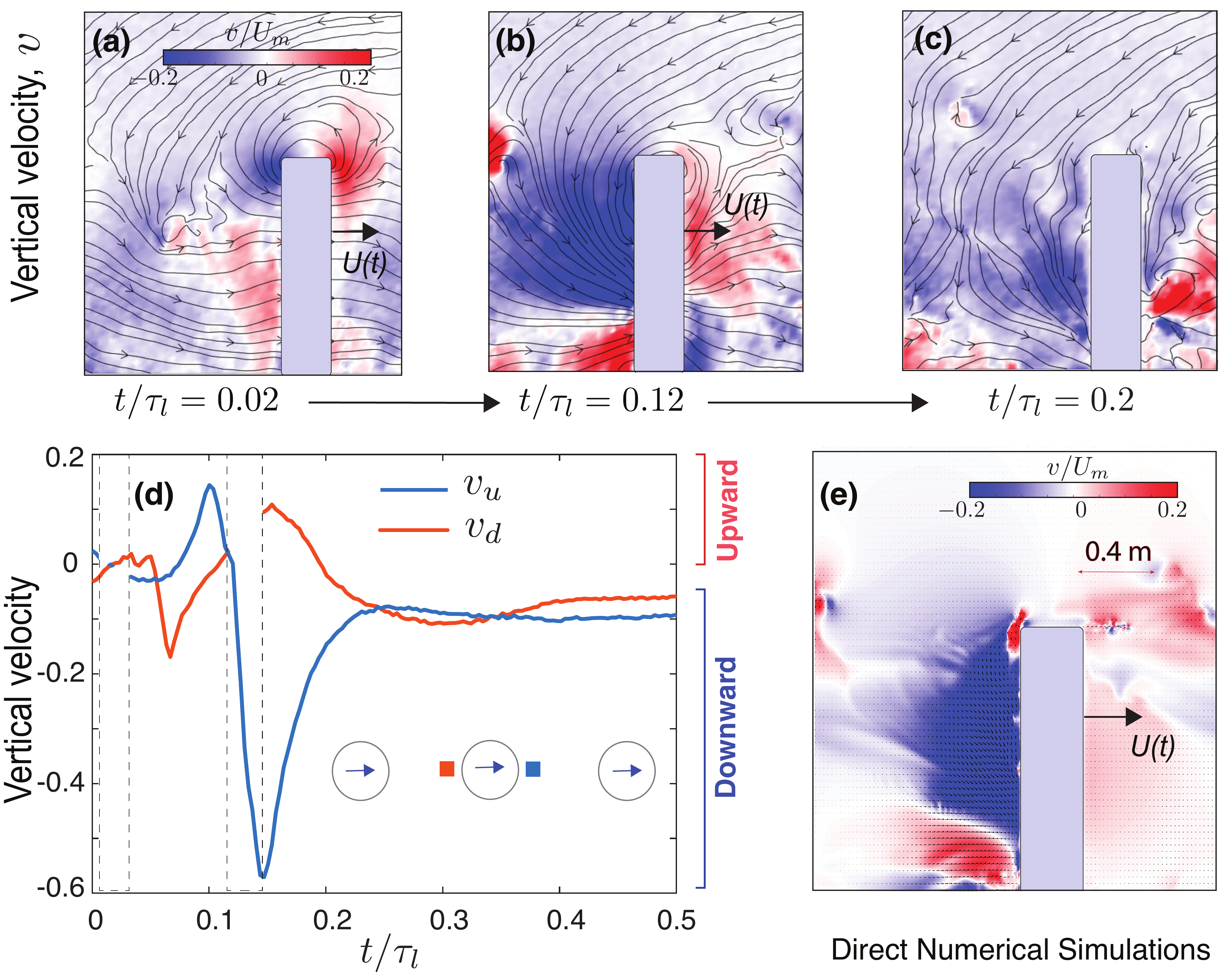}
 \caption{\footnotesize{Experiments and DNS of the velocity fields and {``downwash''} current generated by the periodic start-stop movements in the waiting line. (a)-(c) Vertical velocity fields obtained from Particle Image Velocimetry (PIV) with streamlines showing the strong {``downwash''} generated soon after the cylinders are in motion. Vertical velocity at the blue and red marked locations in front of and behind the cylinder are shown in (d).
 The blanked intervals in (d) correspond to the crossing of the cylinder.
  In (e), the vertical velocity field obtained from DNS, at $t/\tau_l = 0.12$, is shown. The flow patterns are similar and the normalized velocity is comparable in magnitude to the experimentally obtained PIV field.}}
 \label{fig_DNS_PIV_vertical_velocity}
 \vspace{-0.2 cm}
\end{figure}

The simulation domain is of size $9.3D_c$ in the walking direction (horizontal), $5.3D_c$ in the transverse direction (horizontal), and $8.8D_c$ in the vertical direction, which accommodates three cylinders between the periodic domains. The moving cylinders are maintained to have a similar height-to-diameter ratio as in the experiments, and the Reynolds number was set to $Re = 6000$. This value of Re, although lower than in the experiment, is sufficiently large that the flow patterns are nearly Reynolds-independent (shown later in Fig.~\ref{fig_DNS_PIV_vertical_velocity}). To account for the buoyancy change due to varying ambient temperature range of $24-35^\circ$C, $Gr$ was varied in the range of range [$0 - 2.5] \times 10^7$. Here, $Gr = 0$ ($\hat{U}=0$) corresponds to the condition with zero thermal gradients (an idealization), and $Gr = 2.5 \times 10^7$ ($\hat{U}=0.83$) corresponds to a temperature difference between breath and ambient of $\Delta T \approx 4 ^\circ$C following Refs. \cite{yang2022increased,houdas2013human}.

The spreading of airborne particles (dye) is studied using a standard advection-diffusion equation for a scalar $C$ \cite{Diffusion}:
\begin{equation}
 \begin{aligned}
  \frac{\partial C}{\partial t}+{\bf u} \cdot\nabla C= Sc^{-1} Re^{-1} \nabla^2 C + \phi_C.
 \end{aligned}
\end{equation}

The scalar $C$ can be taken as as either a temperature field or as a dye concentration, with a Schmidt number $Sc=D/\nu$, and with $D$ the diffusion coefficient of the scalar. We set the scalar diffusivity to be equal to the fluid kinematic viscosity. This reduces resolution requirements while reproducing the main features of the flow that the scalar can reveal \cite{mckeown2018cascade,mckeown2020turbulence}. For the small to moderate temperature differences, common in natural convection, thermal effects scale linearly as a buoyancy force \cite{zeytounian2003joseph}.

For the particle concentration analysis, a spatial resolution of $256 \times 384 \times 768$ was used, with clustering of points in the $y$-direction close to the bottom wall. The flow was initialized for zero velocity, and the cylinders were moved for four start-stop cycles. After this, the dye field is initialized to zero everywhere, except for a spatial region around the ``infected'' cylinder, mimicking the dye release in the experiment (Fig.~\ref{figure2_dye_DNS_currentsettling}a). Additionally, one thousand Lagrangian tracer particles distributed were released from this region. We simulated the system for four more cycles, collecting velocities, scalar fields, and particle trajectories.

\begin{figure*}[!tbp]
\centering
\includegraphics[width=0.93\linewidth]{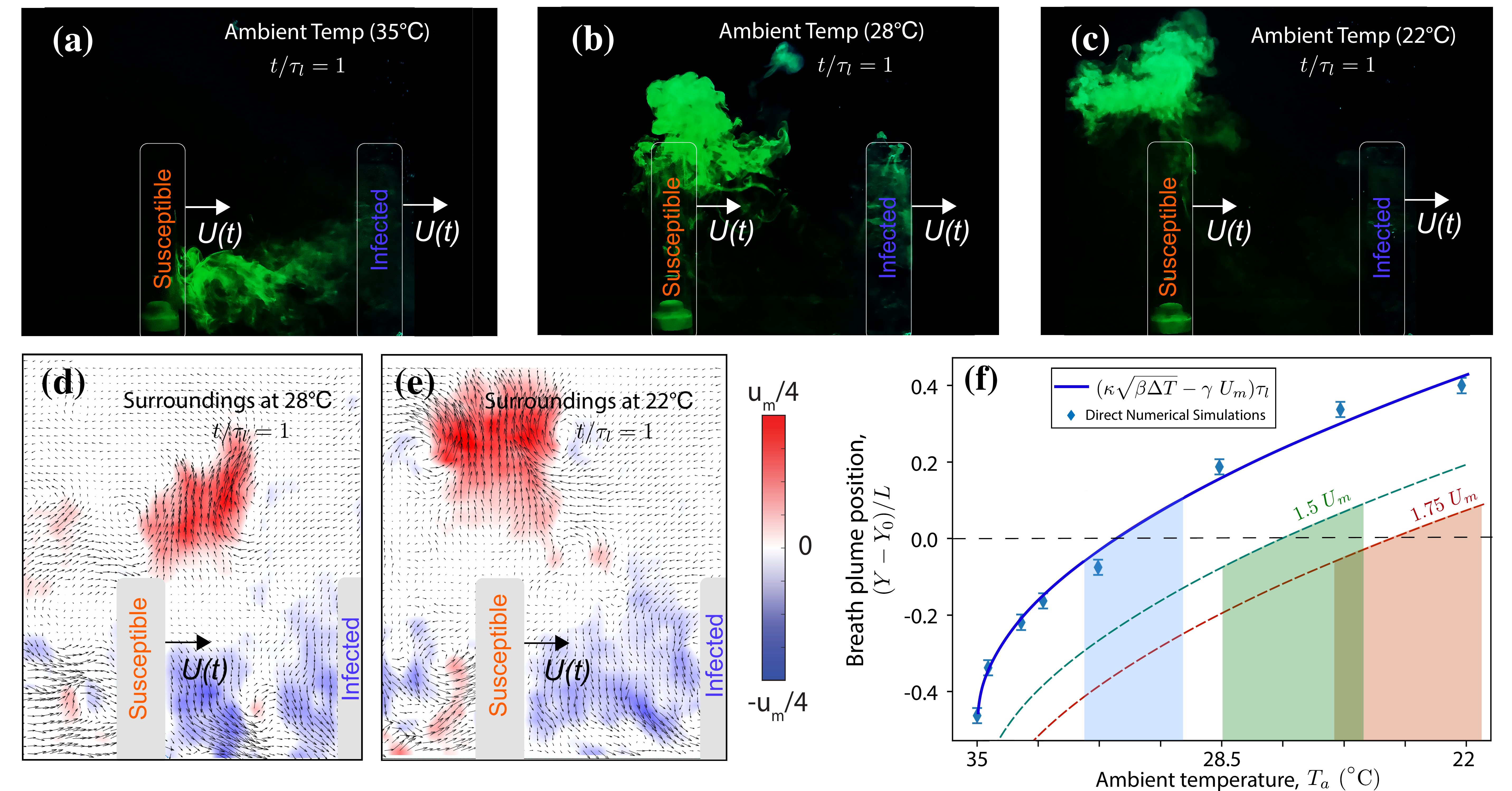}
\caption{\footnotesize{Counteracting roles of line kinematics and plume buoyancy on the spread of airborne particles. In (a) is shown a visualization snapshot of the location of a colored dye (a proxy for infectious aerosols), originally released from the cylinder on the right side, for a relatively high ambient temperature condition, after one waiting time period, $t/\tau_l =1$. This case is equivalent to an ambient temperature $T_a \approx 35^\circ$C. The released aerosols descend due to the aerodynamic ``downwash'' created by the walking and stopping movement. Cases (a)-(c) show how the same kind of particles end up at different locations under the competing effects of the downwash and the breath plume's buoyancy under different ambient temperature conditions ($T_a \approx 35 ^\circ$C,  $28^\circ$C, and $22^\circ$C in (a), (b), and (c), respectively). 
The competition between buoyancy and line kinematics results leads to non-monotonic trends of cross-infection risk for the susceptible individuals trailing in the line. In (d), (e), the vertical velocity PIV fields show the buoyancy plume at the ambient conditions of (b) and (c), respectively.  The mean vertical position of the breathing plume after one period ($t=\tau_l$)  vs. ambient temperature is shown in (f), along with the predictions of the competing flow model (blue curve). The location of the plume is sensitive to the walking speed, $U_m$, by the predictions given for the green and red curves at 50\% and 75\% increase in walking speed, respectively. The horizontal shaded bands (blue, green, and red) show the temperature range of heightened infection risks corresponding to the walking speeds, 0.5 m/s, 0.75 m/s, and 0.87 m/s.}}
\label{figure5_buoyancy_experiment_heightmodel}
\end{figure*}

\section*{Results}

We first consider the buoyancy-less case ($\hat{U}=0$) with an experiment where a passive, fluorescent dye is released from one of the cylinders, \#2 (the infectious) as shown in Fig.~\ref{figure2_dye_DNS_currentsettling}. The dye can be considered as a proxy for aerosols released by an infectious individual during breathing, and by observing its evolution in time, we can analyze where the infectious aerosols get transported to after one cycle of walking and waiting. The top view and side view recordings are shown in Fig.~\ref{figure2_dye_DNS_currentsettling}.a,b \& c and insets, corresponding to $t/\tau_l = 0, 0.08$ \& $0.5$, respectively. The visualization indicates that the released particles end up right in front of the trailing, susceptible person with no noticeable dilution (as signified by the intensity of the dye in the top view of Fig.~\ref{figure2_dye_DNS_currentsettling}.c). Remarkably, despite the dye being nearly perfectly density-matched with the fluid, it begins to sink in front of the susceptible walker (\#3 cylinder) soon after the walking phase ends.

To understand the origin of the downward drift of the dye, we study the underlying flow field using both time-resolved particle-image velocimetry (PIV) experiments and direct numerical simulations (DNS). Figure \ref{fig_DNS_PIV_vertical_velocity}a-c shows the vertical velocity and the flow streamlines at successive points of the movement. A strong, circulating flow is created initially ($t/\tau_l = 0.02$), which traps the aerosols at their release location, even though the individuals in the line move forward (see also inset to Fig.~\ref{figure2_dye_DNS_currentsettling}b). However, this is immediately followed by a sudden {\it downwash} current (blue regions in Fig.~\ref{fig_DNS_PIV_vertical_velocity}b at $t/\tau_l = 0.02$). This downwash current is responsible for the downward drift of the dye (observed earlier in Fig.~\ref{figure2_dye_DNS_currentsettling}) and is reminiscent of the wake with a "starting vortex" generated behind an airplane wing or that behind a circulation-generating bluff body undergoing a start-stop motion\cite{pullin1980some}. To quantify the strength of the {\it downwash}, in Fig.~\ref{fig_DNS_PIV_vertical_velocity}d we sample the vertical velocity at fixed points in space while the walkers (cylinders) move past these points. A strong downward current is formed downstream of the walker, which peaks in strength ($v \approx -0.6 U_m$) around $t \approx 0.15 \tau_l$. By $t \approx 0.25 \tau_l$, the downdraft velocity has reduced in strength, to about 10\% of the walking speed, but the downdraft persists at this value for the remainder of the waiting period.
From a purely kinematic viewpoint, therefore, a sinking of the infectious plume can be expected (as seen in Fig.~\ref{figure2_dye_DNS_currentsettling}), which presumably reduces the infection risk for the trailing individuals in the waiting line.

We compare the velocity fields from DNS with those measured from PIV (Fig.~\ref{fig_DNS_PIV_vertical_velocity} b \& e) at $t/\tau_l = 0.12$. Despite the difference in Reynolds number in the experiment and DNS, nearly the same large-scale flow features are observed, including the strong downwash. Furthermore, the normalized velocities are remarkably similar in magnitude. These further ascertain the assumption of nearly Reynolds-independent flow fields within the Re range in our study ($[0.5 - 1.8] \times 10^4$). Drawing from these insights, we model the sinking speed of the breath plume as $ \gamma \ U_m$, where the prefactor $\gamma \sim  0.1$ was experimentally (or numerically) confirmed.

\begin{figure*}[!htbp]
\centering
\includegraphics[width=1.00\linewidth]{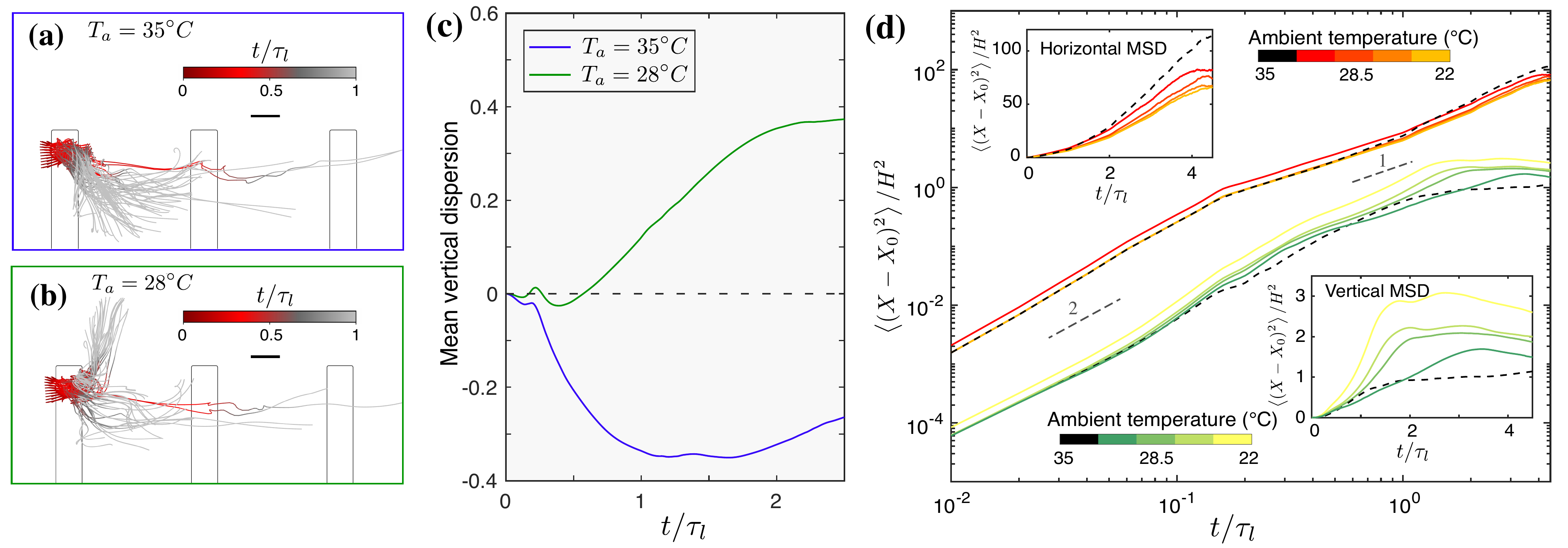}
\caption{Lagrangian trajectories and dispersion statistics of (potentially infectious) particles in the waiting line, obtained from DNS. Plot (a)  and (b), correspond to the non-buoyant case and buoyant case, respectively, with the particles released from the sides of the cylinder, during one period after release. The colors red to grey represent the time after release (see also SI). The mean vertical dispersion of tracers is shown for the non-buoyant (blue) and buoyant (green) cases is shown in (c). Mean squared displacement (MSD) of the Lagrangian tracer particles released from the sides of the cylinders for different net buoyancy conditions are summarized in (d). The upper curves are the horizontal MSD, and the lower lines are the vertical MSD of the tracers on a log-log scale. Insets show the same on data on a linear scale. Here, $\tau_l$ and the movement duration $t/\tau_l = 1/6$ are matched with the experiment.}
\label{figure4_MSD}
\end{figure*}

Because thermal gradients are often present in waiting lines, the downwash will be opposed by the buoyancy of the breathed air, complicating the above picture and modifying the ultimate location of the infectious plume. We systematically vary $\hat{U}$ and study the location where the particles end up after one time period of line movement. The particle displays a remarkable sensitivity on the ambient condition, as visualized using colored dye in Fig.~\ref{figure5_buoyancy_experiment_heightmodel}a,b \& c for ambient temperatures equivalent to 35$^\circ$C, 28$^\circ$C, and 22$^\circ$C, respectively. While in the absence of buoyancy (Fig.~\ref{figure5_buoyancy_experiment_heightmodel}a), the released dye ends up at the base of the cylinders, for high buoyancy (Fig.~\ref{figure5_buoyancy_experiment_heightmodel}c), the dye ends up above the cylinders. In between, at an intermediate buoyancy (Fig.~\ref{figure5_buoyancy_experiment_heightmodel}b) the particles remain in front of the face of the (susceptible) trailing walker, indicating a high cross-infection risk. 
This effect is highlighted when looking at the vertical velocity of the case at {\color{black}28$^\circ$C and 22$^\circ$C} in Fig.~\ref{figure5_buoyancy_experiment_heightmodel}d \& e: aside from a good qualitative comparison between experiment and simulation, we can see how in both cases the infected plume has sufficient buoyancy to oppose the ever-present downwash current from the wake effects in the waiting line. In a more general context, the condition of the highest infection risk would depend crucially on the competition between line kinematics (walking speed) and the ambient temperature (buoyancy). The increased infection risk arising from this upward current can be analyzed through a quantitative model that captures the combined effects of the counter-currents on the spread of infectious aerosols. The cloud of aerosols may be expected to rise at a velocity scale, $ \sim \sqrt{g\beta\Delta T}$ which is counteracted by the downwash velocity of $\gamma U_m$. Here, $\beta$ is the thermal expansivity of the fluid, and $\gamma \approx 0.1$ was obtained previously (see SI~\cite{supp} for further details). The resulting vertical drift after one cycle of line movement can be written as, 

\begin{align}
 Y-Y_0= ({\kappa} \sqrt{\beta \Delta T} - \gamma \ U_m)  \tau_l,
 \label{eq_plume_model}
 \end{align} 
 where $\kappa \sim \sqrt{g  \ell_p}$ and $\ell_p$ is the size scale of the released plume.

Fig.~\ref{figure5_buoyancy_experiment_heightmodel}f shows the model predictions for the vertical drift of the infectious plume for different ambient temperatures and walking speeds. The model shows good agreement with the vertical position of the plume obtained from the simulations (blue symbols). Additionally, we can observe that the highest infection risks occur for an intermediate temperature range. The downward current (second term in Eq.~\ref{eq_plume_model}) grows linearly with the walking speed, $U_m$. As discussed earlier, the typical walking speed in impeded indoor spaces could vary in the range [0.5 - 0.87] m/s. Thus, for a slightly faster walking speed (given by the green and red curves in Fig.~\ref{figure5_buoyancy_experiment_heightmodel}f), the downwash current would gain in intensity.  We extend the model to the higher walking speeds (green curve at $1.5\times$, and red curve at $1.75\times$). It becomes apparent that the risks can be elevated for ambient temperature ranges of  $25 -28.5^\circ$C and $21.5-25.5^\circ$C, as shown by the horizontal shaded areas (green and red shaded areas, respectively). In summary, a broad range of intermediate ambient temperatures -- within the human comfort temperature range -- could potentially heighten the infection risks in indoor waiting lines. In contrast, our analyses show that colder and warmer ambients both suppress the risks.

Using the data from the simulations, we analyze the dispersion of particles released around the cylinder during the walking-waiting cycle. In Fig.~\ref{figure4_MSD}a-b, we show how the trajectory of particles for two cases at $T_a =  35$ and $28^\circ$ C. The particle transport behavior depends crucially on the ambient temperature: a relatively small change in $T_a$ ($\sim 4^\circ$C) results in significantly different particle trajectories  ($\hat{U}$ varies from $0-0.59$). The effect is quantified in Fig.~\ref{figure4_MSD}c, which shows the mean vertical displacement for the same two ambient temperature cases presented in Fig.~\ref{figure4_MSD}a,b. Whether the particles end up near the {\it zero-line} or not depends on both the walking speed and the temperature gradient. Next, we show the mean-squared-displacement (MSD) in the horizontal direction (Fig.~\ref{figure4_MSD}) for ambient temperatures $T_a$ in the range $22-35^\circ$C, resulting in a $Gr$ variation of $[0 - 2.5] \times 10^7$. Note that the time axis is normalized with the period of line movements ($\tilde{t} = t/\tau_l$). Despite the large variation in $Gr$, the scaling and transition times of the MSDs are not affected significantly. We observe a clear ballistic range ($\sim t^2$) that extends up to $\tilde{t}\sim 10^{-1}$, followed by a gradual approach to diffusive scaling ($\sim t$), suggesting that the line kinematics dominate the dispersion scaling. Upon further inspection, this transition time scale of ballistic-to-diffusive scaling coincides with the walking time scale $\tau_m$. The vertical dispersion plateaus around $\tilde{t} \sim 3$, as the dispersion distance approaches the size (height) of the room. The variations caused by buoyancy are still evident when the MSDs are plotted on a linear scale (see insets). Ambient temperature has opposing effects on horizontal and vertical dispersion of particles (see upper and lower insets) --  whereas a higher ambient temperature increases vertical dispersion, it suppresses horizontal dispersion. This can be rationalized by returning to the Lagrangian trajectories of the particles shown in Fig.~\ref{figure4_MSD}a,b. In the absence of buoyancy, the particles are pulled downwards by the underlying downwash current while also being carried forward (horizontally) by the mean drift around the walkers. However, when the ambient temperature is increased, the horizontal dispersion is suppressed due to the lifting of plumes above the height of the currents created by the walkers. Note that the particle transport discussed here is for small aerosolized droplets ($\leq 10~ \mu$m), where the droplet Stokes number and Froude number are sufficiently small that we can safely ignore gravitational settling and inertial clustering of the particles~\cite{mathai2016microbubbles,mathai2020bubbly} for the time scales of interest here. 
Moreover, during regular breathing of individuals in the line, the fraction of large droplets is relatively low, unlike the case of violent respiratory expulsions such as a sneeze or a cough \cite{somsen2020small,bourouiba2020turbulent}.

\section*{Summary and Outlook}
This paper has provided a glimpse into the complex flow patterns created by periodic start-stop motions mimicking the kinematics of everyday interactions in waiting lines around us, from a grocery store or cafe to airports or a voting center. We have revealed the existence of fluid dynamical counter-currents that arise from the competing roles of shear (walking) and buoyancy (thermal gradients between the breath plume and ambient) in a traditional waiting line, by using a combination of laboratory experiments and direct numerical simulations. These counter-currents can cause significant variations in the infection risks for individuals waiting in the line. While faster-moving waiting lines strengthen the downward current ({\it downwash}), a low ambient temperature increases the buoyancy-induced upward draft of the breath plume. The ratio of the walking speed, $U_m$, of individuals to the free-fall velocity scale, $U_b$, of the breath plume determines the conditions of elevated risks.  This is set up when the breath plume is very effectively counteracted by the wake {\it downwash}, causing the aerosols to linger in the air at an infectious height and at minimal dilution levels for extended time (Fig.~\ref{figure5_buoyancy_experiment_heightmodel}b). 
Depending on the walking speed of individuals in the line, infection risks may be heightened at intermediate ambient temperatures ($22 - 30^\circ$C), which fall within the human comfort range \cite{nicol2020range}. However, the risks may be reduced in both hotter ($\geq$ 32 $^\circ$C) and colder ($\leq$ 22 $^\circ$C) ambients, which can have important implications across climatic zones (tropical to temperate climate biomes).  

The findings presented here highlight the stark contrast in the transmission mechanisms dominating ``static'' social interactions, where the flows generated the kinematics of individuals are ignored in the modeling. We modeled the breath plume dynamics under the combined influence of the {\it downwash} and the thermal gradient and obtain an expression for the vertical position of the breath plume after one walking/waiting cycle as $Y-Y_0= ({\kappa} \sqrt{\beta \Delta T} - \gamma \ U_m)  \tau_l$. The Lagrangian statistics of the dispersion of particles show that the horizontal dispersion (MSD$_h$) is enhanced (suppressed) at lower (higher) ambient temperatures, while for the vertical MSD the temperature dependence is reversed. 

Our conclusions rely on the average walking speed and the discontinuous motion in the line, which are crucial in creating the periodic {\it downwash} current. Since the walking speed, ambient temperature, and wait-walk time period ratio are all variable, the problem in itself presents a rich complexity in outcomes, which might account for some of the variability in infection patterns recorded in literature \cite{morawska2021physics,mathai2021air}. It should, however, be recognized that the topic of cross-infection risks in a waiting line is truly multivariate, with biological, environmental, and behavioral factors \cite{morawska2021physics}. Our work addresses the effects of line kinematics and temperature in a simplified waiting line setting, assuming a poorly ventilated indoor space. 
Furthermore, our findings should not be extended to the case of continuously moving (walking or running) lines i.e. a reduced frequency $f^* \equiv  \tau_m/\tau_l \to 1$, in which case the shear-driven circulation is expected to take a different form from what we report here. This is outside the scope of the present work. 
Future efforts at mitigation could be targeted at new, modified line arrangements where individuals are not positioned behind each other, or by designing lines in which either of the two identified counter-currents is selectively suppressed, allowing for aerosols to be either pulled up or pushed down. Ventilation approaches specifically aimed for the removal of plumes could also be effective measures to reduce the cross-infection risks in waiting lines.

 \small{{\bf Acknowledgements:} We thank Yuanhang Zhu, Anupam Pandey, Xiaojue Zhu for useful insights into the experiments and numerical simulations.}


\begin{thebibliography}{61}%
\makeatletter
\providecommand \@ifxundefined [1]{%
 \@ifx{#1\undefined}
}%
\providecommand \@ifnum [1]{%
 \ifnum #1\expandafter \@firstoftwo
 \else \expandafter \@secondoftwo
 \fi
}%
\providecommand \@ifx [1]{%
 \ifx #1\expandafter \@firstoftwo
 \else \expandafter \@secondoftwo
 \fi
}%
\providecommand \natexlab [1]{#1}%
\providecommand \enquote  [1]{``#1''}%
\providecommand \bibnamefont  [1]{#1}%
\providecommand \bibfnamefont [1]{#1}%
\providecommand \citenamefont [1]{#1}%
\providecommand \href@noop [0]{\@secondoftwo}%
\providecommand \href [0]{\begingroup \@sanitize@url \@href}%
\providecommand \@href[1]{\@@startlink{#1}\@@href}%
\providecommand \@@href[1]{\endgroup#1\@@endlink}%
\providecommand \@sanitize@url [0]{\catcode `\\12\catcode `\$12\catcode
  `\&12\catcode `\#12\catcode `\^12\catcode `\_12\catcode `\%12\relax}%
\providecommand \@@startlink[1]{}%
\providecommand \@@endlink[0]{}%
\providecommand \url  [0]{\begingroup\@sanitize@url \@url }%
\providecommand \@url [1]{\endgroup\@href {#1}{\urlprefix }}%
\providecommand \urlprefix  [0]{URL }%
\providecommand \Eprint [0]{\href }%
\providecommand \doibase [0]{https://doi.org/}%
\providecommand \selectlanguage [0]{\@gobble}%
\providecommand \bibinfo  [0]{\@secondoftwo}%
\providecommand \bibfield  [0]{\@secondoftwo}%
\providecommand \translation [1]{[#1]}%
\providecommand \BibitemOpen [0]{}%
\providecommand \bibitemStop [0]{}%
\providecommand \bibitemNoStop [0]{.\EOS\space}%
\providecommand \EOS [0]{\spacefactor3000\relax}%
\providecommand \BibitemShut  [1]{\csname bibitem#1\endcsname}%
\let\auto@bib@innerbib\@empty
\bibitem [{\citenamefont {Bourouiba}(2021)}]{bourouiba2021fluid}%
  \BibitemOpen
  \bibfield  {author} {\bibinfo {author} {\bibfnamefont {L.}~\bibnamefont
  {Bourouiba}},\ }\href@noop {} {\bibfield  {journal} {\bibinfo  {journal}
  {Annual Review of Fluid Mechanics}\ }\textbf {\bibinfo {volume} {53}},\
  \bibinfo {pages} {473} (\bibinfo {year} {2021})}\BibitemShut {NoStop}%
\bibitem [{\citenamefont {Abkarian}\ \emph {et~al.}(2020)\citenamefont
  {Abkarian}, \citenamefont {Mendez}, \citenamefont {Xue}, \citenamefont
  {Yang},\ and\ \citenamefont {Stone}}]{abkarian2020speech}%
  \BibitemOpen
  \bibfield  {author} {\bibinfo {author} {\bibfnamefont {M.}~\bibnamefont
  {Abkarian}}, \bibinfo {author} {\bibfnamefont {S.}~\bibnamefont {Mendez}},
  \bibinfo {author} {\bibfnamefont {N.}~\bibnamefont {Xue}}, \bibinfo {author}
  {\bibfnamefont {F.}~\bibnamefont {Yang}},\ and\ \bibinfo {author}
  {\bibfnamefont {H.~A.}\ \bibnamefont {Stone}},\ }\href@noop {} {\bibfield
  {journal} {\bibinfo  {journal} {Proceedings of the National Academy of
  Sciences}\ }\textbf {\bibinfo {volume} {117}},\ \bibinfo {pages} {25237}
  (\bibinfo {year} {2020})}\BibitemShut {NoStop}%
\bibitem [{\citenamefont {P{\"o}hlker}\ \emph {et~al.}(2021)\citenamefont
  {P{\"o}hlker}, \citenamefont {Kr{\"u}ger}, \citenamefont {F{\"o}rster},
  \citenamefont {Berkemeier}, \citenamefont {Elbert}, \citenamefont
  {Fr{\"o}hlich-Nowoisky}, \citenamefont {P{\"o}schl}, \citenamefont
  {P{\"o}hlker}, \citenamefont {Bagheri}, \citenamefont {Bodenschatz} \emph
  {et~al.}}]{pohlker2021respiratory}%
  \BibitemOpen
  \bibfield  {author} {\bibinfo {author} {\bibfnamefont {M.~L.}\ \bibnamefont
  {P{\"o}hlker}}, \bibinfo {author} {\bibfnamefont {O.~O.}\ \bibnamefont
  {Kr{\"u}ger}}, \bibinfo {author} {\bibfnamefont {J.-D.}\ \bibnamefont
  {F{\"o}rster}}, \bibinfo {author} {\bibfnamefont {T.}~\bibnamefont
  {Berkemeier}}, \bibinfo {author} {\bibfnamefont {W.}~\bibnamefont {Elbert}},
  \bibinfo {author} {\bibfnamefont {J.}~\bibnamefont {Fr{\"o}hlich-Nowoisky}},
  \bibinfo {author} {\bibfnamefont {U.}~\bibnamefont {P{\"o}schl}}, \bibinfo
  {author} {\bibfnamefont {C.}~\bibnamefont {P{\"o}hlker}}, \bibinfo {author}
  {\bibfnamefont {G.}~\bibnamefont {Bagheri}}, \bibinfo {author} {\bibfnamefont
  {E.}~\bibnamefont {Bodenschatz}}, \emph {et~al.},\ }\href@noop {} {\bibfield
  {journal} {\bibinfo  {journal} {arXiv preprint arXiv:2103.01188}\ } (\bibinfo
  {year} {2021})}\BibitemShut {NoStop}%
\bibitem [{\citenamefont {Chong}\ \emph {et~al.}(2021)\citenamefont {Chong},
  \citenamefont {Ng}, \citenamefont {Hori}, \citenamefont {Yang}, \citenamefont
  {Verzicco},\ and\ \citenamefont {Lohse}}]{chong2021extended}%
  \BibitemOpen
  \bibfield  {author} {\bibinfo {author} {\bibfnamefont {K.~L.}\ \bibnamefont
  {Chong}}, \bibinfo {author} {\bibfnamefont {C.~S.}\ \bibnamefont {Ng}},
  \bibinfo {author} {\bibfnamefont {N.}~\bibnamefont {Hori}}, \bibinfo {author}
  {\bibfnamefont {R.}~\bibnamefont {Yang}}, \bibinfo {author} {\bibfnamefont
  {R.}~\bibnamefont {Verzicco}},\ and\ \bibinfo {author} {\bibfnamefont
  {D.}~\bibnamefont {Lohse}},\ }\href@noop {} {\bibfield  {journal} {\bibinfo
  {journal} {Physical review letters}\ }\textbf {\bibinfo {volume} {126}},\
  \bibinfo {pages} {034502} (\bibinfo {year} {2021})}\BibitemShut {NoStop}%
\bibitem [{\citenamefont {Bourouiba}(2020)}]{bourouiba2020turbulent}%
  \BibitemOpen
  \bibfield  {author} {\bibinfo {author} {\bibfnamefont {L.}~\bibnamefont
  {Bourouiba}},\ }\href@noop {} {\bibfield  {journal} {\bibinfo  {journal}
  {Jama}\ }\textbf {\bibinfo {volume} {323}},\ \bibinfo {pages} {1837}
  (\bibinfo {year} {2020})}\BibitemShut {NoStop}%
\bibitem [{\citenamefont {Mittal}\ \emph {et~al.}(2023)\citenamefont {Mittal},
  \citenamefont {Breuer},\ and\ \citenamefont {Seo}}]{mittal2023flow}%
  \BibitemOpen
  \bibfield  {author} {\bibinfo {author} {\bibfnamefont {R.}~\bibnamefont
  {Mittal}}, \bibinfo {author} {\bibfnamefont {K.}~\bibnamefont {Breuer}},\
  and\ \bibinfo {author} {\bibfnamefont {J.~H.}\ \bibnamefont {Seo}},\
  }\href@noop {} {\bibfield  {journal} {\bibinfo  {journal} {Annual Review of
  Fluid Mechanics}\ }\textbf {\bibinfo {volume} {55}} (\bibinfo {year}
  {2023})}\BibitemShut {NoStop}%
\bibitem [{\citenamefont {Tang}\ \emph {et~al.}(2011)\citenamefont {Tang},
  \citenamefont {Noakes}, \citenamefont {Nielsen}, \citenamefont {Eames},
  \citenamefont {Nicolle}, \citenamefont {Li},\ and\ \citenamefont
  {Settles}}]{Tang2011}%
  \BibitemOpen
  \bibfield  {author} {\bibinfo {author} {\bibfnamefont {J.}~\bibnamefont
  {Tang}}, \bibinfo {author} {\bibfnamefont {C.}~\bibnamefont {Noakes}},
  \bibinfo {author} {\bibfnamefont {P.}~\bibnamefont {Nielsen}}, \bibinfo
  {author} {\bibfnamefont {I.}~\bibnamefont {Eames}}, \bibinfo {author}
  {\bibfnamefont {A.}~\bibnamefont {Nicolle}}, \bibinfo {author} {\bibfnamefont
  {Y.}~\bibnamefont {Li}},\ and\ \bibinfo {author} {\bibfnamefont
  {G.}~\bibnamefont {Settles}},\ }\href
  {https://doi.org/https://doi.org/10.1016/j.jhin.2010.09.037} {\bibfield
  {journal} {\bibinfo  {journal} {J. Hosp. Infec.}\ }\textbf {\bibinfo {volume}
  {77}},\ \bibinfo {pages} {213 } (\bibinfo {year} {2011})}\BibitemShut
  {NoStop}%
\bibitem [{\citenamefont {Lai}\ \emph {et~al.}(2012)\citenamefont {Lai},
  \citenamefont {Poon},\ and\ \citenamefont {Cheung}}]{Lai2012}%
  \BibitemOpen
  \bibfield  {author} {\bibinfo {author} {\bibfnamefont {A.~C.~K.}\
  \bibnamefont {Lai}}, \bibinfo {author} {\bibfnamefont {C.~K.~M.}\
  \bibnamefont {Poon}},\ and\ \bibinfo {author} {\bibfnamefont {A.~C.~T.}\
  \bibnamefont {Cheung}},\ }\href {https://doi.org/10.1098/rsif.2011.0537}
  {\bibfield  {journal} {\bibinfo  {journal} {J. R. Soc. Interface}\ }\textbf
  {\bibinfo {volume} {9}},\ \bibinfo {pages} {938} (\bibinfo {year}
  {2012})}\BibitemShut {NoStop}%
\bibitem [{\citenamefont {Greenhalgh}\ \emph {et~al.}(2020)\citenamefont
  {Greenhalgh}, \citenamefont {Schmid}, \citenamefont {Czypionka},
  \citenamefont {Bassler},\ and\ \citenamefont {Gruer}}]{Greenhalgh2020}%
  \BibitemOpen
  \bibfield  {author} {\bibinfo {author} {\bibfnamefont {T.}~\bibnamefont
  {Greenhalgh}}, \bibinfo {author} {\bibfnamefont {M.~B.}\ \bibnamefont
  {Schmid}}, \bibinfo {author} {\bibfnamefont {T.}~\bibnamefont {Czypionka}},
  \bibinfo {author} {\bibfnamefont {D.}~\bibnamefont {Bassler}},\ and\ \bibinfo
  {author} {\bibfnamefont {L.}~\bibnamefont {Gruer}},\ }\href@noop {}
  {\bibfield  {journal} {\bibinfo  {journal} {BMJ}\ }\textbf {\bibinfo {volume}
  {369}} (\bibinfo {year} {2020})}\BibitemShut {NoStop}%
\bibitem [{\citenamefont {Leung}\ \emph {et~al.}(2020)\citenamefont {Leung},
  \citenamefont {Chu}, \citenamefont {Shiu}, \citenamefont {Chan},\ and\
  \citenamefont {McDevitt~et al.}}]{Leung2020}%
  \BibitemOpen
  \bibfield  {author} {\bibinfo {author} {\bibfnamefont {N.~H.~L.}\
  \bibnamefont {Leung}}, \bibinfo {author} {\bibfnamefont {D.~K.~W.}\
  \bibnamefont {Chu}}, \bibinfo {author} {\bibfnamefont {E.~Y.~C.}\
  \bibnamefont {Shiu}}, \bibinfo {author} {\bibfnamefont {K.-H.}\ \bibnamefont
  {Chan}},\ and\ \bibinfo {author} {\bibfnamefont {J.~J.}\ \bibnamefont
  {McDevitt~et al.}},\ }\href {https://doi.org/10.1038/s41591-020-0843-2}
  {\bibfield  {journal} {\bibinfo  {journal} {Nat. Med.}\ }\textbf {\bibinfo
  {volume} {26}},\ \bibinfo {pages} {676} (\bibinfo {year} {2020})}\BibitemShut
  {NoStop}%
\bibitem [{\citenamefont {Lee}\ \emph {et~al.}(2008)\citenamefont {Lee},
  \citenamefont {Grinshpun},\ and\ \citenamefont
  {Reponen}}]{lee2008respiratory}%
  \BibitemOpen
  \bibfield  {author} {\bibinfo {author} {\bibfnamefont {S.-A.}\ \bibnamefont
  {Lee}}, \bibinfo {author} {\bibfnamefont {S.~A.}\ \bibnamefont {Grinshpun}},\
  and\ \bibinfo {author} {\bibfnamefont {T.}~\bibnamefont {Reponen}},\
  }\href@noop {} {\bibfield  {journal} {\bibinfo  {journal} {Ann. Occup. Hyg.}\
  }\textbf {\bibinfo {volume} {52}},\ \bibinfo {pages} {177} (\bibinfo {year}
  {2008})}\BibitemShut {NoStop}%
\bibitem [{\citenamefont {Povaiah}(2020)}]{plasticpanel2020roshun}%
  \BibitemOpen
  \bibfield  {author} {\bibinfo {author} {\bibfnamefont {R.}~\bibnamefont
  {Povaiah}},\ }\href@noop {} {\bibfield  {journal} {\bibinfo  {journal} {The
  Quint}\ }\textbf {\bibinfo {volume} {{1}}},\ \bibinfo {pages} {9} (\bibinfo
  {year} {2020})}\BibitemShut {NoStop}%
\bibitem [{\citenamefont {Morawska}\ \emph {et~al.}(2020)\citenamefont
  {Morawska}, \citenamefont {Tang}, \citenamefont {Bahnfleth}, \citenamefont
  {Bluyssen}, \citenamefont {Boerstra},\ and\ \citenamefont
  {et~al.}}]{Morawska2020}%
  \BibitemOpen
  \bibfield  {author} {\bibinfo {author} {\bibfnamefont {L.}~\bibnamefont
  {Morawska}}, \bibinfo {author} {\bibfnamefont {J.~W.}\ \bibnamefont {Tang}},
  \bibinfo {author} {\bibfnamefont {W.}~\bibnamefont {Bahnfleth}}, \bibinfo
  {author} {\bibfnamefont {P.~M.}\ \bibnamefont {Bluyssen}}, \bibinfo {author}
  {\bibfnamefont {A.}~\bibnamefont {Boerstra}},\ and\ \bibinfo {author}
  {\bibfnamefont {G.~B.}\ \bibnamefont {et~al.}},\ }\href
  {https://doi.org/https://doi.org/10.1016/j.envint.2020.105832} {\bibfield
  {journal} {\bibinfo  {journal} {Environ. Int.}\ }\textbf {\bibinfo {volume}
  {142}},\ \bibinfo {pages} {105832} (\bibinfo {year} {2020})}\BibitemShut
  {NoStop}%
\bibitem [{\citenamefont {Zhang}\ \emph {et~al.}(2020)\citenamefont {Zhang},
  \citenamefont {Li}, \citenamefont {Zhang}, \citenamefont {Wang},\ and\
  \citenamefont {Molina}}]{zhang2020identifying}%
  \BibitemOpen
  \bibfield  {author} {\bibinfo {author} {\bibfnamefont {R.}~\bibnamefont
  {Zhang}}, \bibinfo {author} {\bibfnamefont {Y.}~\bibnamefont {Li}}, \bibinfo
  {author} {\bibfnamefont {A.~L.}\ \bibnamefont {Zhang}}, \bibinfo {author}
  {\bibfnamefont {Y.}~\bibnamefont {Wang}},\ and\ \bibinfo {author}
  {\bibfnamefont {M.~J.}\ \bibnamefont {Molina}},\ }\href@noop {} {\bibfield
  {journal} {\bibinfo  {journal} {Proc. Natl. Acad. Sci.}\ }\textbf {\bibinfo
  {volume} {117}},\ \bibinfo {pages} {14857} (\bibinfo {year}
  {2020})}\BibitemShut {NoStop}%
\bibitem [{\citenamefont {Yu}\ \emph {et~al.}(2004)\citenamefont {Yu},
  \citenamefont {Li}, \citenamefont {Wong}, \citenamefont {Tam}, \citenamefont
  {Chan}, \citenamefont {Lee}, \citenamefont {Leung},\ and\ \citenamefont
  {Ho}}]{yu2004evidence}%
  \BibitemOpen
  \bibfield  {author} {\bibinfo {author} {\bibfnamefont {I.~T.}\ \bibnamefont
  {Yu}}, \bibinfo {author} {\bibfnamefont {Y.}~\bibnamefont {Li}}, \bibinfo
  {author} {\bibfnamefont {T.~W.}\ \bibnamefont {Wong}}, \bibinfo {author}
  {\bibfnamefont {W.}~\bibnamefont {Tam}}, \bibinfo {author} {\bibfnamefont
  {A.~T.}\ \bibnamefont {Chan}}, \bibinfo {author} {\bibfnamefont {J.~H.}\
  \bibnamefont {Lee}}, \bibinfo {author} {\bibfnamefont {D.~Y.}\ \bibnamefont
  {Leung}},\ and\ \bibinfo {author} {\bibfnamefont {T.}~\bibnamefont {Ho}},\
  }\href@noop {} {\bibfield  {journal} {\bibinfo  {journal} {N. Engl. J. Med.}\
  }\textbf {\bibinfo {volume} {350}},\ \bibinfo {pages} {1731} (\bibinfo {year}
  {2004})}\BibitemShut {NoStop}%
\bibitem [{\citenamefont {Abkarian}\ and\ \citenamefont
  {Stone}(2020)}]{abkarian2020stretching}%
  \BibitemOpen
  \bibfield  {author} {\bibinfo {author} {\bibfnamefont {M.}~\bibnamefont
  {Abkarian}}\ and\ \bibinfo {author} {\bibfnamefont {H.~A.}\ \bibnamefont
  {Stone}},\ }\href@noop {} {\bibfield  {journal} {\bibinfo  {journal}
  {Physical Review Fluids}\ }\textbf {\bibinfo {volume} {5}},\ \bibinfo {pages}
  {102301} (\bibinfo {year} {2020})}\BibitemShut {NoStop}%
\bibitem [{\citenamefont {He}\ \emph {et~al.}(2021)\citenamefont {He},
  \citenamefont {Liu}, \citenamefont {Elson}, \citenamefont {Vogt},
  \citenamefont {Maranville},\ and\ \citenamefont {Hong}}]{he2021airborne}%
  \BibitemOpen
  \bibfield  {author} {\bibinfo {author} {\bibfnamefont {R.}~\bibnamefont
  {He}}, \bibinfo {author} {\bibfnamefont {W.}~\bibnamefont {Liu}}, \bibinfo
  {author} {\bibfnamefont {J.}~\bibnamefont {Elson}}, \bibinfo {author}
  {\bibfnamefont {R.}~\bibnamefont {Vogt}}, \bibinfo {author} {\bibfnamefont
  {C.}~\bibnamefont {Maranville}},\ and\ \bibinfo {author} {\bibfnamefont
  {J.}~\bibnamefont {Hong}},\ }\href@noop {} {\bibfield  {journal} {\bibinfo
  {journal} {Physics of Fluids}\ }\textbf {\bibinfo {volume} {33}},\ \bibinfo
  {pages} {057107} (\bibinfo {year} {2021})}\BibitemShut {NoStop}%
\bibitem [{\citenamefont {Bhagat}\ \emph {et~al.}(2020)\citenamefont {Bhagat},
  \citenamefont {Wykes}, \citenamefont {Dalziel},\ and\ \citenamefont
  {Linden}}]{bhagat2020effects}%
  \BibitemOpen
  \bibfield  {author} {\bibinfo {author} {\bibfnamefont {R.~K.}\ \bibnamefont
  {Bhagat}}, \bibinfo {author} {\bibfnamefont {M.~D.}\ \bibnamefont {Wykes}},
  \bibinfo {author} {\bibfnamefont {S.~B.}\ \bibnamefont {Dalziel}},\ and\
  \bibinfo {author} {\bibfnamefont {P.}~\bibnamefont {Linden}},\ }\href@noop {}
  {\bibfield  {journal} {\bibinfo  {journal} {Journal of Fluid Mechanics}\
  }\textbf {\bibinfo {volume} {903}},\ \bibinfo {pages} {F1} (\bibinfo {year}
  {2020})}\BibitemShut {NoStop}%
\bibitem [{\citenamefont {Gupta}\ \emph {et~al.}(2010)\citenamefont {Gupta},
  \citenamefont {Lin},\ and\ \citenamefont {Chen}}]{Gupta2010}%
  \BibitemOpen
  \bibfield  {author} {\bibinfo {author} {\bibfnamefont {J.~K.}\ \bibnamefont
  {Gupta}}, \bibinfo {author} {\bibfnamefont {C.-H.}\ \bibnamefont {Lin}},\
  and\ \bibinfo {author} {\bibfnamefont {Q.}~\bibnamefont {Chen}},\ }\href
  {https://doi.org/10.1111/j.1600-0668.2009.00623.x} {\bibfield  {journal}
  {\bibinfo  {journal} {Indoor Air}\ }\textbf {\bibinfo {volume} {20}},\
  \bibinfo {pages} {31} (\bibinfo {year} {2010})}\BibitemShut {NoStop}%
\bibitem [{\citenamefont {Meselson}(2020)}]{Meselson2020}%
  \BibitemOpen
  \bibfield  {author} {\bibinfo {author} {\bibfnamefont {M.}~\bibnamefont
  {Meselson}},\ }\href@noop {} {\bibfield  {journal} {\bibinfo  {journal} {N.
  Engl. J. Med}\ }\textbf {\bibinfo {volume} {382}},\ \bibinfo {pages} {2063}
  (\bibinfo {year} {2020})}\BibitemShut {NoStop}%
\bibitem [{\citenamefont {Yan}\ \emph {et~al.}(2018)\citenamefont {Yan},
  \citenamefont {Grantham}, \citenamefont {Pantelic}, \citenamefont
  {De~Mesquita}, \citenamefont {Albert}, \citenamefont {Liu}, \citenamefont
  {Ehrman}, \citenamefont {Milton}, \citenamefont {Consortium} \emph
  {et~al.}}]{yan2018infectious}%
  \BibitemOpen
  \bibfield  {author} {\bibinfo {author} {\bibfnamefont {J.}~\bibnamefont
  {Yan}}, \bibinfo {author} {\bibfnamefont {M.}~\bibnamefont {Grantham}},
  \bibinfo {author} {\bibfnamefont {J.}~\bibnamefont {Pantelic}}, \bibinfo
  {author} {\bibfnamefont {P.~J.~B.}\ \bibnamefont {De~Mesquita}}, \bibinfo
  {author} {\bibfnamefont {B.}~\bibnamefont {Albert}}, \bibinfo {author}
  {\bibfnamefont {F.}~\bibnamefont {Liu}}, \bibinfo {author} {\bibfnamefont
  {S.}~\bibnamefont {Ehrman}}, \bibinfo {author} {\bibfnamefont {D.~K.}\
  \bibnamefont {Milton}}, \bibinfo {author} {\bibfnamefont {E.}~\bibnamefont
  {Consortium}}, \emph {et~al.},\ }\href@noop {} {\bibfield  {journal}
  {\bibinfo  {journal} {Proc. Natl. Acad. Sci.}\ }\textbf {\bibinfo {volume}
  {115}},\ \bibinfo {pages} {1081} (\bibinfo {year} {2018})}\BibitemShut
  {NoStop}%
\bibitem [{\citenamefont {W{\"o}lfel}\ \emph {et~al.}(2020)\citenamefont
  {W{\"o}lfel}, \citenamefont {Corman}, \citenamefont {Guggemos}, \citenamefont
  {Seilmaier},\ and\ \citenamefont {Zange~et al.}}]{wolfel2020virological}%
  \BibitemOpen
  \bibfield  {author} {\bibinfo {author} {\bibfnamefont {R.}~\bibnamefont
  {W{\"o}lfel}}, \bibinfo {author} {\bibfnamefont {V.~M.}\ \bibnamefont
  {Corman}}, \bibinfo {author} {\bibfnamefont {W.}~\bibnamefont {Guggemos}},
  \bibinfo {author} {\bibfnamefont {M.}~\bibnamefont {Seilmaier}},\ and\
  \bibinfo {author} {\bibfnamefont {S.}~\bibnamefont {Zange~et al.}},\
  }\href@noop {} {\bibfield  {journal} {\bibinfo  {journal} {Nature}\ }\textbf
  {\bibinfo {volume} {581}},\ \bibinfo {pages} {465} (\bibinfo {year}
  {2020})}\BibitemShut {NoStop}%
\bibitem [{\citenamefont {Yang}\ \emph {et~al.}(2011)\citenamefont {Yang},
  \citenamefont {Elankumaran},\ and\ \citenamefont
  {Marr}}]{yang2011concentrations}%
  \BibitemOpen
  \bibfield  {author} {\bibinfo {author} {\bibfnamefont {W.}~\bibnamefont
  {Yang}}, \bibinfo {author} {\bibfnamefont {S.}~\bibnamefont {Elankumaran}},\
  and\ \bibinfo {author} {\bibfnamefont {L.~C.}\ \bibnamefont {Marr}},\
  }\href@noop {} {\bibfield  {journal} {\bibinfo  {journal} {J. R. Soc.
  Interface}\ }\textbf {\bibinfo {volume} {8}},\ \bibinfo {pages} {1176}
  (\bibinfo {year} {2011})}\BibitemShut {NoStop}%
\bibitem [{\citenamefont {Scharfman}\ \emph {et~al.}(2016)\citenamefont
  {Scharfman}, \citenamefont {Techet}, \citenamefont {Bush},\ and\
  \citenamefont {Bourouiba}}]{scharfman2016visualization}%
  \BibitemOpen
  \bibfield  {author} {\bibinfo {author} {\bibfnamefont {B.}~\bibnamefont
  {Scharfman}}, \bibinfo {author} {\bibfnamefont {A.}~\bibnamefont {Techet}},
  \bibinfo {author} {\bibfnamefont {J.}~\bibnamefont {Bush}},\ and\ \bibinfo
  {author} {\bibfnamefont {L.}~\bibnamefont {Bourouiba}},\ }\href@noop {}
  {\bibfield  {journal} {\bibinfo  {journal} {Exp Fluids}\ }\textbf {\bibinfo
  {volume} {57}},\ \bibinfo {pages} {24} (\bibinfo {year} {2016})}\BibitemShut
  {NoStop}%
\bibitem [{\citenamefont {Bahl}\ \emph {et~al.}(2020)\citenamefont {Bahl},
  \citenamefont {Doolan}, \citenamefont {de~Silva}, \citenamefont {Chughtai},
  \citenamefont {Bourouiba},\ and\ \citenamefont
  {MacIntyre}}]{bahl2020airborne}%
  \BibitemOpen
  \bibfield  {author} {\bibinfo {author} {\bibfnamefont {P.}~\bibnamefont
  {Bahl}}, \bibinfo {author} {\bibfnamefont {C.}~\bibnamefont {Doolan}},
  \bibinfo {author} {\bibfnamefont {C.}~\bibnamefont {de~Silva}}, \bibinfo
  {author} {\bibfnamefont {A.~A.}\ \bibnamefont {Chughtai}}, \bibinfo {author}
  {\bibfnamefont {L.}~\bibnamefont {Bourouiba}},\ and\ \bibinfo {author}
  {\bibfnamefont {C.~R.}\ \bibnamefont {MacIntyre}},\ }\href@noop {} {\bibfield
   {journal} {\bibinfo  {journal} {J. Infect. Dis}\ }\textbf {\bibinfo {volume}
  {189}},\ \bibinfo {pages} {1093} (\bibinfo {year} {2020})}\BibitemShut
  {NoStop}%
\bibitem [{\citenamefont {Bourouiba}\ \emph {et~al.}(2014)\citenamefont
  {Bourouiba}, \citenamefont {Dehandschoewercker},\ and\ \citenamefont
  {Bush}}]{bourouiba2014violent}%
  \BibitemOpen
  \bibfield  {author} {\bibinfo {author} {\bibfnamefont {L.}~\bibnamefont
  {Bourouiba}}, \bibinfo {author} {\bibfnamefont {E.}~\bibnamefont
  {Dehandschoewercker}},\ and\ \bibinfo {author} {\bibfnamefont {J.~W.}\
  \bibnamefont {Bush}},\ }\href@noop {} {\bibfield  {journal} {\bibinfo
  {journal} {J. Fluid Mech.}\ }\textbf {\bibinfo {volume} {745}},\ \bibinfo
  {pages} {537} (\bibinfo {year} {2014})}\BibitemShut {NoStop}%
\bibitem [{\citenamefont {Van~Doremalen}\ \emph {et~al.}(2020)\citenamefont
  {Van~Doremalen}, \citenamefont {Bushmaker}, \citenamefont {Morris},
  \citenamefont {Holbrook},\ and\ \citenamefont {Gamble~et
  al.}}]{van2020aerosol}%
  \BibitemOpen
  \bibfield  {author} {\bibinfo {author} {\bibfnamefont {N.}~\bibnamefont
  {Van~Doremalen}}, \bibinfo {author} {\bibfnamefont {T.}~\bibnamefont
  {Bushmaker}}, \bibinfo {author} {\bibfnamefont {D.~H.}\ \bibnamefont
  {Morris}}, \bibinfo {author} {\bibfnamefont {M.~G.}\ \bibnamefont
  {Holbrook}},\ and\ \bibinfo {author} {\bibfnamefont {A.}~\bibnamefont
  {Gamble~et al.}},\ }\href@noop {} {\bibfield  {journal} {\bibinfo  {journal}
  {N. Engl. J. Med}\ }\textbf {\bibinfo {volume} {382}},\ \bibinfo {pages}
  {1564} (\bibinfo {year} {2020})}\BibitemShut {NoStop}%
\bibitem [{\citenamefont {Stadnytskyi}\ \emph {et~al.}(2020)\citenamefont
  {Stadnytskyi}, \citenamefont {Bax}, \citenamefont {Bax},\ and\ \citenamefont
  {Anfinrud}}]{stadnytskyi2020airborne}%
  \BibitemOpen
  \bibfield  {author} {\bibinfo {author} {\bibfnamefont {V.}~\bibnamefont
  {Stadnytskyi}}, \bibinfo {author} {\bibfnamefont {C.~E.}\ \bibnamefont
  {Bax}}, \bibinfo {author} {\bibfnamefont {A.}~\bibnamefont {Bax}},\ and\
  \bibinfo {author} {\bibfnamefont {P.}~\bibnamefont {Anfinrud}},\ }\href@noop
  {} {\bibfield  {journal} {\bibinfo  {journal} {Proc. Natl. Acad. Sci.}\
  }\textbf {\bibinfo {volume} {117}},\ \bibinfo {pages} {11875} (\bibinfo
  {year} {2020})}\BibitemShut {NoStop}%
\bibitem [{\citenamefont {Mathai}\ \emph {et~al.}(2022)\citenamefont {Mathai},
  \citenamefont {Das},\ and\ \citenamefont {Breuer}}]{mathai2022aerosol}%
  \BibitemOpen
  \bibfield  {author} {\bibinfo {author} {\bibfnamefont {V.}~\bibnamefont
  {Mathai}}, \bibinfo {author} {\bibfnamefont {A.}~\bibnamefont {Das}},\ and\
  \bibinfo {author} {\bibfnamefont {K.}~\bibnamefont {Breuer}},\ }\href@noop {}
  {\bibfield  {journal} {\bibinfo  {journal} {Physics of Fluids}\ }\textbf
  {\bibinfo {volume} {34}},\ \bibinfo {pages} {021904} (\bibinfo {year}
  {2022})}\BibitemShut {NoStop}%
\bibitem [{\citenamefont {Yang}\ \emph {et~al.}(2020)\citenamefont {Yang},
  \citenamefont {Pahlavan}, \citenamefont {Mendez}, \citenamefont {Abkarian},\
  and\ \citenamefont {Stone}}]{yang2020towards}%
  \BibitemOpen
  \bibfield  {author} {\bibinfo {author} {\bibfnamefont {F.}~\bibnamefont
  {Yang}}, \bibinfo {author} {\bibfnamefont {A.~A.}\ \bibnamefont {Pahlavan}},
  \bibinfo {author} {\bibfnamefont {S.}~\bibnamefont {Mendez}}, \bibinfo
  {author} {\bibfnamefont {M.}~\bibnamefont {Abkarian}},\ and\ \bibinfo
  {author} {\bibfnamefont {H.~A.}\ \bibnamefont {Stone}},\ }\href@noop {}
  {\bibfield  {journal} {\bibinfo  {journal} {Physical Review Fluids}\ }\textbf
  {\bibinfo {volume} {5}},\ \bibinfo {pages} {122501} (\bibinfo {year}
  {2020})}\BibitemShut {NoStop}%
\bibitem [{\citenamefont {Beggs}\ \emph {et~al.}(2010)\citenamefont {Beggs},
  \citenamefont {Shepherd},\ and\ \citenamefont {Kerr}}]{beggs2010potential}%
  \BibitemOpen
  \bibfield  {author} {\bibinfo {author} {\bibfnamefont {C.~B.}\ \bibnamefont
  {Beggs}}, \bibinfo {author} {\bibfnamefont {S.~J.}\ \bibnamefont
  {Shepherd}},\ and\ \bibinfo {author} {\bibfnamefont {K.~G.}\ \bibnamefont
  {Kerr}},\ }\href@noop {} {\bibfield  {journal} {\bibinfo  {journal} {BMC
  infectious diseases}\ }\textbf {\bibinfo {volume} {10}},\ \bibinfo {pages}
  {1} (\bibinfo {year} {2010})}\BibitemShut {NoStop}%
\bibitem [{\citenamefont {Mathai}(2021)}]{mathai2021air}%
  \BibitemOpen
  \bibfield  {author} {\bibinfo {author} {\bibfnamefont {V.}~\bibnamefont
  {Mathai}},\ }\href@noop {} {\bibfield  {journal} {\bibinfo  {journal}
  {Physics Today}\ }\textbf {\bibinfo {volume} {74}},\ \bibinfo {pages} {66}
  (\bibinfo {year} {2021})}\BibitemShut {NoStop}%
\bibitem [{\citenamefont {Tang}\ \emph {et~al.}(2009)\citenamefont {Tang},
  \citenamefont {Liebner}, \citenamefont {Craven},\ and\ \citenamefont
  {Settles}}]{tang2009schlieren}%
  \BibitemOpen
  \bibfield  {author} {\bibinfo {author} {\bibfnamefont {J.~W.}\ \bibnamefont
  {Tang}}, \bibinfo {author} {\bibfnamefont {T.~J.}\ \bibnamefont {Liebner}},
  \bibinfo {author} {\bibfnamefont {B.~A.}\ \bibnamefont {Craven}},\ and\
  \bibinfo {author} {\bibfnamefont {G.~S.}\ \bibnamefont {Settles}},\
  }\href@noop {} {\bibfield  {journal} {\bibinfo  {journal} {Journal of the
  Royal Society Interface}\ }\textbf {\bibinfo {volume} {6}},\ \bibinfo {pages}
  {S727} (\bibinfo {year} {2009})}\BibitemShut {NoStop}%
\bibitem [{\citenamefont {Mittal}\ \emph {et~al.}(2020)\citenamefont {Mittal},
  \citenamefont {Ni},\ and\ \citenamefont {Seo}}]{mittal2020flow}%
  \BibitemOpen
  \bibfield  {author} {\bibinfo {author} {\bibfnamefont {R.}~\bibnamefont
  {Mittal}}, \bibinfo {author} {\bibfnamefont {R.}~\bibnamefont {Ni}},\ and\
  \bibinfo {author} {\bibfnamefont {J.-H.}\ \bibnamefont {Seo}},\ }\href@noop
  {} {\bibfield  {journal} {\bibinfo  {journal} {J. Fluid Mech.}\ }\textbf
  {\bibinfo {volume} {894}},\ \bibinfo {pages} {330} (\bibinfo {year}
  {2020})}\BibitemShut {NoStop}%
\bibitem [{\citenamefont {Supplemental}(2024)}]{supp}%
  \BibitemOpen
  \bibfield  {author} {\bibinfo {author} {\bibnamefont {Supplemental}},\
  }\href@noop {} {\bibfield  {journal} {\bibinfo  {journal} {Material}\ }
  (\bibinfo {year} {2024})}\BibitemShut {NoStop}%
\bibitem [{\citenamefont {Niemela}\ \emph {et~al.}(2000)\citenamefont
  {Niemela}, \citenamefont {Skrbek}, \citenamefont {Sreenivasan},\ and\
  \citenamefont {Donnelly}}]{niemela2000turbulent}%
  \BibitemOpen
  \bibfield  {author} {\bibinfo {author} {\bibfnamefont {J.}~\bibnamefont
  {Niemela}}, \bibinfo {author} {\bibfnamefont {L.}~\bibnamefont {Skrbek}},
  \bibinfo {author} {\bibfnamefont {K.}~\bibnamefont {Sreenivasan}},\ and\
  \bibinfo {author} {\bibfnamefont {R.}~\bibnamefont {Donnelly}},\ }\href@noop
  {} {\bibfield  {journal} {\bibinfo  {journal} {Nature}\ }\textbf {\bibinfo
  {volume} {404}},\ \bibinfo {pages} {837} (\bibinfo {year}
  {2000})}\BibitemShut {NoStop}%
\bibitem [{\citenamefont {Ahlers}\ \emph {et~al.}(2009)\citenamefont {Ahlers},
  \citenamefont {Grossmann},\ and\ \citenamefont {Lohse}}]{ahlers2009heat}%
  \BibitemOpen
  \bibfield  {author} {\bibinfo {author} {\bibfnamefont {G.}~\bibnamefont
  {Ahlers}}, \bibinfo {author} {\bibfnamefont {S.}~\bibnamefont {Grossmann}},\
  and\ \bibinfo {author} {\bibfnamefont {D.}~\bibnamefont {Lohse}},\
  }\href@noop {} {\bibfield  {journal} {\bibinfo  {journal} {Reviews of modern
  physics}\ }\textbf {\bibinfo {volume} {81}},\ \bibinfo {pages} {503}
  (\bibinfo {year} {2009})}\BibitemShut {NoStop}%
\bibitem [{\citenamefont {Iyer}\ \emph {et~al.}(2020)\citenamefont {Iyer},
  \citenamefont {Scheel}, \citenamefont {Schumacher},\ and\ \citenamefont
  {Sreenivasan}}]{iyer2020classical}%
  \BibitemOpen
  \bibfield  {author} {\bibinfo {author} {\bibfnamefont {K.~P.}\ \bibnamefont
  {Iyer}}, \bibinfo {author} {\bibfnamefont {J.~D.}\ \bibnamefont {Scheel}},
  \bibinfo {author} {\bibfnamefont {J.}~\bibnamefont {Schumacher}},\ and\
  \bibinfo {author} {\bibfnamefont {K.~R.}\ \bibnamefont {Sreenivasan}},\
  }\href@noop {} {\bibfield  {journal} {\bibinfo  {journal} {Proceedings of the
  National Academy of Sciences}\ }\textbf {\bibinfo {volume} {117}},\ \bibinfo
  {pages} {7594} (\bibinfo {year} {2020})}\BibitemShut {NoStop}%
\bibitem [{\citenamefont {Lohse}\ and\ \citenamefont
  {Xia}(2010)}]{lohse2010small}%
  \BibitemOpen
  \bibfield  {author} {\bibinfo {author} {\bibfnamefont {D.}~\bibnamefont
  {Lohse}}\ and\ \bibinfo {author} {\bibfnamefont {K.-Q.}\ \bibnamefont
  {Xia}},\ }\href@noop {} {\bibfield  {journal} {\bibinfo  {journal} {Annual
  Review of Fluid Mechanics}\ }\textbf {\bibinfo {volume} {42}},\ \bibinfo
  {pages} {335} (\bibinfo {year} {2010})}\BibitemShut {NoStop}%
\bibitem [{\citenamefont {Xia}(2013)}]{xia2013current}%
  \BibitemOpen
  \bibfield  {author} {\bibinfo {author} {\bibfnamefont {K.-Q.}\ \bibnamefont
  {Xia}},\ }\href@noop {} {\bibfield  {journal} {\bibinfo  {journal}
  {Theoretical and Applied Mechanics Letters}\ }\textbf {\bibinfo {volume}
  {3}},\ \bibinfo {pages} {052001} (\bibinfo {year} {2013})}\BibitemShut
  {NoStop}%
\bibitem [{\citenamefont {Chaigne}\ \emph {et~al.}(2023)\citenamefont
  {Chaigne}, \citenamefont {Berhanu},\ and\ \citenamefont
  {Kudrolli}}]{chaigne2023dissolution}%
  \BibitemOpen
  \bibfield  {author} {\bibinfo {author} {\bibfnamefont {M.}~\bibnamefont
  {Chaigne}}, \bibinfo {author} {\bibfnamefont {M.}~\bibnamefont {Berhanu}},\
  and\ \bibinfo {author} {\bibfnamefont {A.}~\bibnamefont {Kudrolli}},\
  }\href@noop {} {\bibfield  {journal} {\bibinfo  {journal} {Proceedings of the
  National Academy of Sciences}\ }\textbf {\bibinfo {volume} {120}},\ \bibinfo
  {pages} {e2301947120} (\bibinfo {year} {2023})}\BibitemShut {NoStop}%
\bibitem [{\citenamefont {Wang}\ \emph {et~al.}(2014)\citenamefont {Wang},
  \citenamefont {Yang}, \citenamefont {Chen}, \citenamefont {Liu},
  \citenamefont {Gruteser},\ and\ \citenamefont {Martin}}]{wang2014tracking}%
  \BibitemOpen
  \bibfield  {author} {\bibinfo {author} {\bibfnamefont {Y.}~\bibnamefont
  {Wang}}, \bibinfo {author} {\bibfnamefont {J.}~\bibnamefont {Yang}}, \bibinfo
  {author} {\bibfnamefont {Y.}~\bibnamefont {Chen}}, \bibinfo {author}
  {\bibfnamefont {H.}~\bibnamefont {Liu}}, \bibinfo {author} {\bibfnamefont
  {M.}~\bibnamefont {Gruteser}},\ and\ \bibinfo {author} {\bibfnamefont
  {R.~P.}\ \bibnamefont {Martin}},\ }in\ \href@noop {} {\emph {\bibinfo
  {booktitle} {Proceedings of the 12th annual international conference on
  Mobile systems, applications, and services}}}\ (\bibinfo {year} {2014})\ pp.\
  \bibinfo {pages} {42--54}\BibitemShut {NoStop}%
\bibitem [{\citenamefont {Donelan}\ and\ \citenamefont
  {Kram}(2000)}]{donelan2000exploring}%
  \BibitemOpen
  \bibfield  {author} {\bibinfo {author} {\bibfnamefont {J.~M.}\ \bibnamefont
  {Donelan}}\ and\ \bibinfo {author} {\bibfnamefont {R.}~\bibnamefont {Kram}},\
  }\href@noop {} {\bibfield  {journal} {\bibinfo  {journal} {Journal of
  Experimental Biology}\ }\textbf {\bibinfo {volume} {203}},\ \bibinfo {pages}
  {2405} (\bibinfo {year} {2000})}\BibitemShut {NoStop}%
\bibitem [{\citenamefont {Alm{\'e}ras}\ \emph {et~al.}(2017)\citenamefont
  {Alm{\'e}ras}, \citenamefont {Mathai}, \citenamefont {Lohse},\ and\
  \citenamefont {Sun}}]{almeras2017experimental}%
  \BibitemOpen
  \bibfield  {author} {\bibinfo {author} {\bibfnamefont {E.}~\bibnamefont
  {Alm{\'e}ras}}, \bibinfo {author} {\bibfnamefont {V.}~\bibnamefont {Mathai}},
  \bibinfo {author} {\bibfnamefont {D.}~\bibnamefont {Lohse}},\ and\ \bibinfo
  {author} {\bibfnamefont {C.}~\bibnamefont {Sun}},\ }\href@noop {} {\bibfield
  {journal} {\bibinfo  {journal} {Journal of fluid mechanics}\ }\textbf
  {\bibinfo {volume} {825}},\ \bibinfo {pages} {1091} (\bibinfo {year}
  {2017})}\BibitemShut {NoStop}%
\bibitem [{\citenamefont {Yang}\ \emph {et~al.}(2022)\citenamefont {Yang},
  \citenamefont {Ng}, \citenamefont {Chong}, \citenamefont {Verzicco},\ and\
  \citenamefont {Lohse}}]{yang2022increased}%
  \BibitemOpen
  \bibfield  {author} {\bibinfo {author} {\bibfnamefont {R.}~\bibnamefont
  {Yang}}, \bibinfo {author} {\bibfnamefont {C.~S.}\ \bibnamefont {Ng}},
  \bibinfo {author} {\bibfnamefont {K.~L.}\ \bibnamefont {Chong}}, \bibinfo
  {author} {\bibfnamefont {R.}~\bibnamefont {Verzicco}},\ and\ \bibinfo
  {author} {\bibfnamefont {D.}~\bibnamefont {Lohse}},\ }\href@noop {}
  {\bibfield  {journal} {\bibinfo  {journal} {Journal of fluid mechanics}\
  }\textbf {\bibinfo {volume} {932}} (\bibinfo {year} {2022})}\BibitemShut
  {NoStop}%
\bibitem [{\citenamefont {Houdas}\ and\ \citenamefont
  {Ring}(2013)}]{houdas2013human}%
  \BibitemOpen
  \bibfield  {author} {\bibinfo {author} {\bibfnamefont {Y.}~\bibnamefont
  {Houdas}}\ and\ \bibinfo {author} {\bibfnamefont {E.}~\bibnamefont {Ring}},\
  }\href@noop {} {\emph {\bibinfo {title} {Human body temperature: its
  measurement and regulation}}}\ (\bibinfo  {publisher} {Springer Science \&
  Business Media},\ \bibinfo {year} {2013})\BibitemShut {NoStop}%
\bibitem [{\citenamefont {Fadlun}\ \emph {et~al.}(2000)\citenamefont {Fadlun},
  \citenamefont {Verzicco}, \citenamefont {Orlandi},\ and\ \citenamefont
  {Mohd-Yusof}}]{fadlun2000combined}%
  \BibitemOpen
  \bibfield  {author} {\bibinfo {author} {\bibfnamefont {E.}~\bibnamefont
  {Fadlun}}, \bibinfo {author} {\bibfnamefont {R.}~\bibnamefont {Verzicco}},
  \bibinfo {author} {\bibfnamefont {P.}~\bibnamefont {Orlandi}},\ and\ \bibinfo
  {author} {\bibfnamefont {J.}~\bibnamefont {Mohd-Yusof}},\ }\href@noop {}
  {\bibfield  {journal} {\bibinfo  {journal} {Journal of computational
  physics}\ }\textbf {\bibinfo {volume} {161}},\ \bibinfo {pages} {35}
  (\bibinfo {year} {2000})}\BibitemShut {NoStop}%
\bibitem [{\citenamefont {Van Der~Poel}\ \emph {et~al.}(2015)\citenamefont {Van
  Der~Poel}, \citenamefont {Ostilla-M{\'o}nico}, \citenamefont {Donners},\ and\
  \citenamefont {Verzicco}}]{van2015pencil}%
  \BibitemOpen
  \bibfield  {author} {\bibinfo {author} {\bibfnamefont {E.~P.}\ \bibnamefont
  {Van Der~Poel}}, \bibinfo {author} {\bibfnamefont {R.}~\bibnamefont
  {Ostilla-M{\'o}nico}}, \bibinfo {author} {\bibfnamefont {J.}~\bibnamefont
  {Donners}},\ and\ \bibinfo {author} {\bibfnamefont {R.}~\bibnamefont
  {Verzicco}},\ }\href@noop {} {\bibfield  {journal} {\bibinfo  {journal}
  {Computers \& Fluids}\ }\textbf {\bibinfo {volume} {116}},\ \bibinfo {pages}
  {10} (\bibinfo {year} {2015})}\BibitemShut {NoStop}%
\bibitem [{\citenamefont {Blass}\ \emph {et~al.}(2020)\citenamefont {Blass},
  \citenamefont {Zhu}, \citenamefont {Verzicco}, \citenamefont {Lohse},\ and\
  \citenamefont {Stevens}}]{blass2020flow}%
  \BibitemOpen
  \bibfield  {author} {\bibinfo {author} {\bibfnamefont {A.}~\bibnamefont
  {Blass}}, \bibinfo {author} {\bibfnamefont {X.}~\bibnamefont {Zhu}}, \bibinfo
  {author} {\bibfnamefont {R.}~\bibnamefont {Verzicco}}, \bibinfo {author}
  {\bibfnamefont {D.}~\bibnamefont {Lohse}},\ and\ \bibinfo {author}
  {\bibfnamefont {R.~J.}\ \bibnamefont {Stevens}},\ }\href@noop {} {\bibfield
  {journal} {\bibinfo  {journal} {Journal of fluid mechanics}\ }\textbf
  {\bibinfo {volume} {897}},\ \bibinfo {pages} {A22} (\bibinfo {year}
  {2020})}\BibitemShut {NoStop}%
\bibitem [{\citenamefont {Blass}\ \emph {et~al.}(2021)\citenamefont {Blass},
  \citenamefont {Verzicco}, \citenamefont {Lohse}, \citenamefont {Stevens},\
  and\ \citenamefont {Krug}}]{blass2021flow}%
  \BibitemOpen
  \bibfield  {author} {\bibinfo {author} {\bibfnamefont {A.}~\bibnamefont
  {Blass}}, \bibinfo {author} {\bibfnamefont {R.}~\bibnamefont {Verzicco}},
  \bibinfo {author} {\bibfnamefont {D.}~\bibnamefont {Lohse}}, \bibinfo
  {author} {\bibfnamefont {R.~J.}\ \bibnamefont {Stevens}},\ and\ \bibinfo
  {author} {\bibfnamefont {D.}~\bibnamefont {Krug}},\ }\href@noop {} {\bibfield
   {journal} {\bibinfo  {journal} {Journal of Fluid Mechanics}\ }\textbf
  {\bibinfo {volume} {906}},\ \bibinfo {pages} {A26} (\bibinfo {year}
  {2021})}\BibitemShut {NoStop}%
\bibitem [{\citenamefont {Jin}\ \emph {et~al.}(2022)\citenamefont {Jin},
  \citenamefont {Wu}, \citenamefont {Zhang}, \citenamefont {Liu},\ and\
  \citenamefont {Zhou}}]{jin2022shear}%
  \BibitemOpen
  \bibfield  {author} {\bibinfo {author} {\bibfnamefont {T.-C.}\ \bibnamefont
  {Jin}}, \bibinfo {author} {\bibfnamefont {J.-Z.}\ \bibnamefont {Wu}},
  \bibinfo {author} {\bibfnamefont {Y.-Z.}\ \bibnamefont {Zhang}}, \bibinfo
  {author} {\bibfnamefont {Y.-L.}\ \bibnamefont {Liu}},\ and\ \bibinfo {author}
  {\bibfnamefont {Q.}~\bibnamefont {Zhou}},\ }\href@noop {} {\bibfield
  {journal} {\bibinfo  {journal} {Journal of Fluid Mechanics}\ }\textbf
  {\bibinfo {volume} {936}},\ \bibinfo {pages} {A28} (\bibinfo {year}
  {2022})}\BibitemShut {NoStop}%
\bibitem [{\citenamefont {Scott A.~Socolofsky}(2005)}]{Diffusion}%
  \BibitemOpen
  \bibfield  {author} {\bibinfo {author} {\bibfnamefont {G.~H.~J.}\
  \bibnamefont {Scott A.~Socolofsky}},\ }\href@noop {} {\bibfield  {journal}
  {\bibinfo  {journal} {xxx}\ } (\bibinfo {year} {2005})}\BibitemShut {NoStop}%
\bibitem [{\citenamefont {McKeown}\ \emph {et~al.}(2018)\citenamefont
  {McKeown}, \citenamefont {Ostilla-M{\'o}nico}, \citenamefont {Pumir},
  \citenamefont {Brenner},\ and\ \citenamefont
  {Rubinstein}}]{mckeown2018cascade}%
  \BibitemOpen
  \bibfield  {author} {\bibinfo {author} {\bibfnamefont {R.}~\bibnamefont
  {McKeown}}, \bibinfo {author} {\bibfnamefont {R.}~\bibnamefont
  {Ostilla-M{\'o}nico}}, \bibinfo {author} {\bibfnamefont {A.}~\bibnamefont
  {Pumir}}, \bibinfo {author} {\bibfnamefont {M.~P.}\ \bibnamefont {Brenner}},\
  and\ \bibinfo {author} {\bibfnamefont {S.~M.}\ \bibnamefont {Rubinstein}},\
  }\href@noop {} {\bibfield  {journal} {\bibinfo  {journal} {Physical Review
  Fluids}\ }\textbf {\bibinfo {volume} {3}},\ \bibinfo {pages} {124702}
  (\bibinfo {year} {2018})}\BibitemShut {NoStop}%
\bibitem [{\citenamefont {McKeown}\ \emph {et~al.}(2020)\citenamefont
  {McKeown}, \citenamefont {Ostilla-M{\'o}nico}, \citenamefont {Pumir},
  \citenamefont {Brenner},\ and\ \citenamefont
  {Rubinstein}}]{mckeown2020turbulence}%
  \BibitemOpen
  \bibfield  {author} {\bibinfo {author} {\bibfnamefont {R.}~\bibnamefont
  {McKeown}}, \bibinfo {author} {\bibfnamefont {R.}~\bibnamefont
  {Ostilla-M{\'o}nico}}, \bibinfo {author} {\bibfnamefont {A.}~\bibnamefont
  {Pumir}}, \bibinfo {author} {\bibfnamefont {M.~P.}\ \bibnamefont {Brenner}},\
  and\ \bibinfo {author} {\bibfnamefont {S.~M.}\ \bibnamefont {Rubinstein}},\
  }\href@noop {} {\bibfield  {journal} {\bibinfo  {journal} {Science advances}\
  }\textbf {\bibinfo {volume} {6}},\ \bibinfo {pages} {eaaz2717} (\bibinfo
  {year} {2020})}\BibitemShut {NoStop}%
\bibitem [{\citenamefont {Zeytounian}(2003)}]{zeytounian2003joseph}%
  \BibitemOpen
  \bibfield  {author} {\bibinfo {author} {\bibfnamefont {R.~K.}\ \bibnamefont
  {Zeytounian}},\ }\href@noop {} {\bibfield  {journal} {\bibinfo  {journal}
  {Comptes Rendus Mecanique}\ }\textbf {\bibinfo {volume} {331}},\ \bibinfo
  {pages} {575} (\bibinfo {year} {2003})}\BibitemShut {NoStop}%
\bibitem [{\citenamefont {Pullin}\ and\ \citenamefont
  {Perry}(1980)}]{pullin1980some}%
  \BibitemOpen
  \bibfield  {author} {\bibinfo {author} {\bibfnamefont {D.}~\bibnamefont
  {Pullin}}\ and\ \bibinfo {author} {\bibfnamefont {A.}~\bibnamefont {Perry}},\
  }\href@noop {} {\bibfield  {journal} {\bibinfo  {journal} {Journal of Fluid
  Mechanics}\ }\textbf {\bibinfo {volume} {97}},\ \bibinfo {pages} {239}
  (\bibinfo {year} {1980})}\BibitemShut {NoStop}%
\bibitem [{\citenamefont {Mathai}\ \emph {et~al.}(2016)\citenamefont {Mathai},
  \citenamefont {Calzavarini}, \citenamefont {Brons}, \citenamefont {Sun},\
  and\ \citenamefont {Lohse}}]{mathai2016microbubbles}%
  \BibitemOpen
  \bibfield  {author} {\bibinfo {author} {\bibfnamefont {V.}~\bibnamefont
  {Mathai}}, \bibinfo {author} {\bibfnamefont {E.}~\bibnamefont {Calzavarini}},
  \bibinfo {author} {\bibfnamefont {J.}~\bibnamefont {Brons}}, \bibinfo
  {author} {\bibfnamefont {C.}~\bibnamefont {Sun}},\ and\ \bibinfo {author}
  {\bibfnamefont {D.}~\bibnamefont {Lohse}},\ }\href@noop {} {\bibfield
  {journal} {\bibinfo  {journal} {Physical Review Letters}\ }\textbf {\bibinfo
  {volume} {117}},\ \bibinfo {pages} {024501} (\bibinfo {year}
  {2016})}\BibitemShut {NoStop}%
\bibitem [{\citenamefont {Mathai}\ \emph {et~al.}(2020)\citenamefont {Mathai},
  \citenamefont {Lohse},\ and\ \citenamefont {Sun}}]{mathai2020bubbly}%
  \BibitemOpen
  \bibfield  {author} {\bibinfo {author} {\bibfnamefont {V.}~\bibnamefont
  {Mathai}}, \bibinfo {author} {\bibfnamefont {D.}~\bibnamefont {Lohse}},\ and\
  \bibinfo {author} {\bibfnamefont {C.}~\bibnamefont {Sun}},\ }\href@noop {}
  {\bibfield  {journal} {\bibinfo  {journal} {Annual Review of Condensed Matter
  Physics}\ }\textbf {\bibinfo {volume} {11}},\ \bibinfo {pages} {529}
  (\bibinfo {year} {2020})}\BibitemShut {NoStop}%
\bibitem [{\citenamefont {Somsen}\ \emph {et~al.}(2020)\citenamefont {Somsen},
  \citenamefont {van Rijn}, \citenamefont {Kooij}, \citenamefont {Bem},\ and\
  \citenamefont {Bonn}}]{somsen2020small}%
  \BibitemOpen
  \bibfield  {author} {\bibinfo {author} {\bibfnamefont {G.~A.}\ \bibnamefont
  {Somsen}}, \bibinfo {author} {\bibfnamefont {C.}~\bibnamefont {van Rijn}},
  \bibinfo {author} {\bibfnamefont {S.}~\bibnamefont {Kooij}}, \bibinfo
  {author} {\bibfnamefont {R.~A.}\ \bibnamefont {Bem}},\ and\ \bibinfo {author}
  {\bibfnamefont {D.}~\bibnamefont {Bonn}},\ }\href@noop {} {\bibfield
  {journal} {\bibinfo  {journal} {The Lancet Respiratory Medicine}\ }\textbf
  {\bibinfo {volume} {8}},\ \bibinfo {pages} {658} (\bibinfo {year}
  {2020})}\BibitemShut {NoStop}%
\bibitem [{\citenamefont {Nicol}\ \emph {et~al.}(2020)\citenamefont {Nicol},
  \citenamefont {Rijal}, \citenamefont {Imagawa},\ and\ \citenamefont
  {Thapa}}]{nicol2020range}%
  \BibitemOpen
  \bibfield  {author} {\bibinfo {author} {\bibfnamefont {F.}~\bibnamefont
  {Nicol}}, \bibinfo {author} {\bibfnamefont {H.~B.}\ \bibnamefont {Rijal}},
  \bibinfo {author} {\bibfnamefont {H.}~\bibnamefont {Imagawa}},\ and\ \bibinfo
  {author} {\bibfnamefont {R.}~\bibnamefont {Thapa}},\ }\href@noop {}
  {\bibfield  {journal} {\bibinfo  {journal} {Energy and Buildings}\ }\textbf
  {\bibinfo {volume} {224}},\ \bibinfo {pages} {110277} (\bibinfo {year}
  {2020})}\BibitemShut {NoStop}%
\bibitem [{\citenamefont {Morawska}\ and\ \citenamefont
  {Buonanno}(2021)}]{morawska2021physics}%
  \BibitemOpen
  \bibfield  {author} {\bibinfo {author} {\bibfnamefont {L.}~\bibnamefont
  {Morawska}}\ and\ \bibinfo {author} {\bibfnamefont {G.}~\bibnamefont
  {Buonanno}},\ }\href@noop {} {\bibfield  {journal} {\bibinfo  {journal}
  {Nature Reviews Physics}\ }\textbf {\bibinfo {volume} {3}},\ \bibinfo {pages}
  {300} (\bibinfo {year} {2021})}\BibitemShut {NoStop}%
\end{thebibliography}
\end{document}